\documentclass[letterpaper,twocolumn,10pt]{article}
\usepackage{usenix2019_v3}

\usepackage{tikz}
\usepackage{amsmath}

\usepackage{amssymb}
\usepackage{eucal}
\usepackage{filecontents}
\usepackage{epsfig,endnotes}
\usepackage{xspace}
\usepackage{amsmath}
\usepackage{xcolor}
\usepackage{pifont}
\usepackage{todonotes}
\usepackage{hyperref}
\usepackage[normalem]{ulem}
\usepackage{balance}
\usepackage{gensymb}
\usepackage{subcaption}
\usepackage{caption}

\usepackage{color,soul}
\sethlcolor{green}

\newcommand{\etal}{\textit{et al}.}
\newcommand{\eg}{e.g., \xspace}
\newcommand{\ie}{i.e., \xspace}
\def\approach{{\sc attack2vec}\xspace}

\soulregister\cite7
\soulregister\ref7
\soulregister\pageref7
\soulregister\approach7
\soulregister\etal7
\soulregister\eg7
\soulregister\ie7
\soulregister\ding0

\author{
{\rm Yun Shen}\\
Symantec Research Labs \\
{\rm \texttt{yun\_shen@symantec.com}}
\and
{\rm Gianluca Stringhini}\\
Boston University \\
{\rm \texttt{gian@bu.edu}}
}

\begin{document}

\date{}

\title{\approach: Leveraging Temporal Word Embeddings to\\ Understand the Evolution of Cyberattacks}

\maketitle

\begin{abstract}

Despite the fact that cyberattacks are constantly growing in complexity, the research community still lacks effective tools to easily monitor and understand them.
  In particular, there is a need for techniques that are able to not only track how prominently certain malicious actions, such as the exploitation of specific vulnerabilities, are exploited in the wild, but also (and more importantly) how these malicious actions factor in as attack steps in more complex cyberattacks.  
In this paper we present \approach, a system that uses temporal word embeddings to model how attack steps are exploited in the wild, and track how they evolve.
  We test \approach on a dataset of billions of security events collected from the customers of a commercial Intrusion Prevention System over a period of two years, and show that our approach is effective in monitoring the emergence of new attack strategies in the wild and in flagging which attack steps are often used together by attackers (e.g., vulnerabilities that are frequently exploited together).
 \approach provides a useful tool for researchers and practitioners to better understand cyberattacks and their evolution, and use this knowledge to improve situational awareness and develop proactive defenses.

\end{abstract}

\section{Introduction}
\label{sec:intro}
Modern cyberattacks have reached high levels of complexity. An attacker who is trying to compromise a computer system has to perform a number of
\emph{attack steps} to achieve her goal~\cite{hutchins2011intelligence}, including reconnaissance (\ie identifying weaknesses on the victim machine), the
actual exploitation, and installing mechanisms to ensure persistence (\eg installing a remote access trojan (RAT) on the
machine~\cite{farinholt2017catch}). Moreover, getting access to the victim machine might not be enough for attackers to achieve what they want, therefore
they might have to perform additional attack steps (\eg exploiting another vulnerability to escalate privileges~\cite{provos2003preventing}). Additionally,
for each of the attack steps that compose the attack, attackers have a choice of executing a variety of malicious actions (\eg exploiting different known
vulnerabilities on the victim system), depending on the exploits that they have available, on the software configuration of the victim machine, and on its
security hygiene (\ie which known vulnerabilities on it have not been patched).

Previous research studied how attack steps (\eg specific Common Vulnerabilities and Exposures (CVEs) being exploited) evolve and are used in isolation~\cite{bilge2012before, nayak2014some, sabottke2015vulnerability}.
While doing so is useful to understand how prominently certain attack steps are exploited in the wild, it does not tell us anything on \emph{how} these attack steps are used as part of complex cyberattacks.
Instead, looking at attack steps in relation to each other can provide researchers and practitioners with invaluable insights into the \emph{modus operandi} of attackers, highlighting important trends in the way attacks are conducted.
In this paper, we define the sequence of attack steps that are commonly performed together with an attack step of interest as its \emph{context}.

Understanding the context in which a vulnerability is exploited in the wild as well as detecting when this context suddenly changes can be very useful for researchers, to better understand the modus operandi of attackers, to improve situational awareness in organizations, and to develop more proactive defenses.
For example, when a new CVE is published, attackers will start attempting to exploit it, and in this process they will first try a number of strategies.
Eventually, once an attacker will succeed in developing an attack that reliably compromises machines, we will observe this strategy being consistently exploited in the wild, potentially because this consolidated attack was commoditized and added to an exploit kit for multiple attackers to use~\cite{grier2012manufacturing}.
This information is useful for defenders, since it allows to design better mitigation strategies that take into account the entire attack, and it can possibly also be used for attack attribution, since the same attacker often uses similar strategies to carry out their attacks~\cite{rid2015attributing}.

However, attack strategies are not stable over time, because new defenses might be deployed that make them ineffective (\eg vulnerabilities getting patched), or simply because the attackers might develop more efficient strategies.
Looking at the context of an attack step (\eg a particular CVE) can help identifying these sudden changes in the way attacks are performed, and prompt proactive defenses.
For example, a number of systems have been proposed that use supervised learning to detect attacks~\cite{bilge2012disclosure,gu2007bothunter,gu2008botminer,sommer2010outside}.
These systems typically need periodic retraining due to the fact that the evolution of attacks over time makes the model that the system was trained on obsolete~\cite{jordaney2017transcend}.
Having a system able to track significant changes in the context associated to a security event could be used to perform a timely retraining of such systems.

To model the context of an attack step, in this paper we adapt techniques that have been proposed in the area of natural language processing.
Word embeddings~\cite{mikolov2013distributed, pennington2014glove} are a powerful tool for modeling relationships between words. 
This technique represents words with low-dimensional vectors based on the surrounding words that appear in the same sentence (\ie the context).
These vectors are able to capture the context of a word and its relationship with the other words, allowing researchers to understand the way in which words are used in various types of language (e.g., on social media~\cite{dhingra2016tweet2vec}).
In a similar way, we can calculate the embedding of an attack step by considering the entire attack sequence as a sentence, and each step as a word.
Upon encoding the relationship between attack steps within the vector space, we can quantitatively study the attack steps appearing in similar contexts in the latent space and understand them in a more meaningful and measurable way. 

As a proxy for the attack steps performed by attackers in the wild, we use the security alerts generated by a commercial Intrusion Prevention System (IPS), collected over a period of two years. Throughout this observation period, we collect 102 snapshots on a weekly basis. Each snapshot contains over 190 million alerts collected from tens of millions unique machines.
Each alert is indicative of the attack step that is performed by an attacker, and our dataset contains over 8k possible alert types, spanning from port scans to exploits for specific CVEs. 
Similar data was used in our previous work, which showed that, although a proxy (\eg they can only monitor attacks for which a detection signature exists), these alerts are useful to study the behavior of attackers in the wild~\cite{shen2018tiresias}.
In the remainder of the paper, we define each alert generated by the IPS a \emph{security event}.

We implement our approach in a system, \approach. Our system takes a stream of security events and computes their context by using temporal word embeddings.
By running \approach on our data, we show that our approach is able to effectively monitor how security events are exploited in the wild.
For example, we can identify when a certain CVE starts getting exploited, when its exploitation becomes stable, and when attackers change strategy in exploiting it. 
By leveraging the similarity between the context of different security events, we can infer which events are often used as part of the same malicious campaign, and this allows us to identify emerging attacks in a more timely manner than the state of the art.
For example, we were able to identify a variant of the Mirai botnet that was scanning the Internet attempting to exploit a CVE relative to Apache Struts, together with IoT-related exploits over 72 weeks before this variant was officially identified.
These findings show that \approach can be an effective tool for researchers and practitioners who need to understand how security events are exploited in the wild and react to sudden changes.

In summary, this paper makes the following contributions:
\begin{itemize}
 \item We show that temporal word embeddings are an effective way to study how attack steps are exploited in the wild and how they evolve. 
 \item We show how \approach can be used to understand the emergence, the evolution, and the characteristics of attack steps in relation to the wider context in which they are exploited.
 \item We discuss how \approach can be effectively used to identify emerging attack campaigns several weeks before they are publicly disclosed.
\end{itemize}

\section{Motivation}
\label{sec:motivation}

\begin{figure*}[t]
	\centering
	\includegraphics[width=\linewidth]{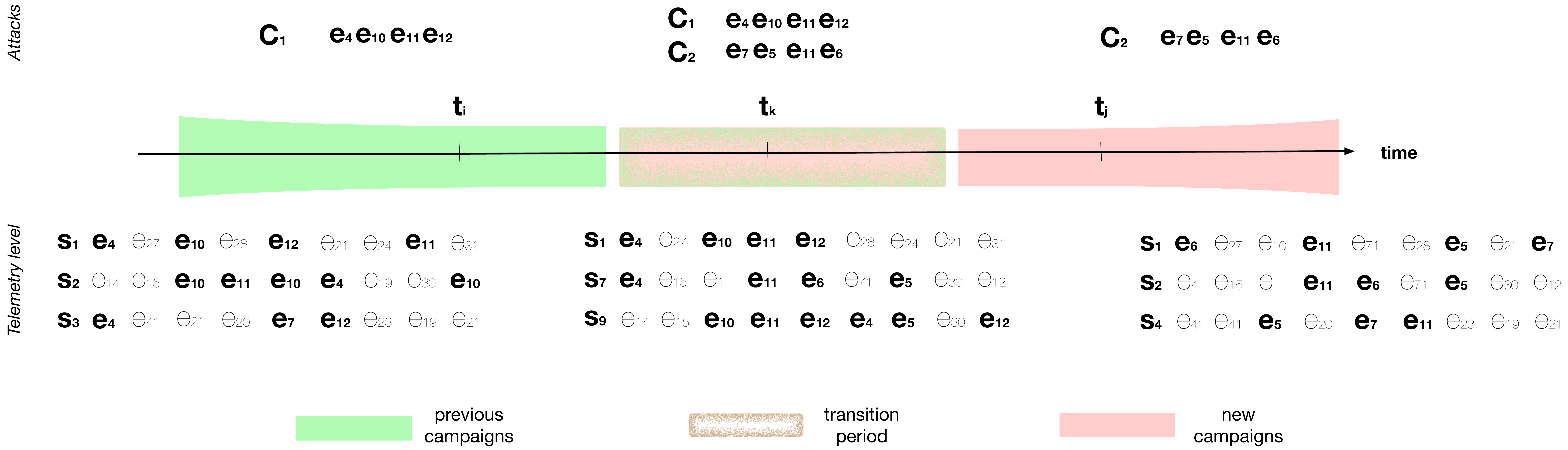}
	\caption{A real-world example of security event evolution. Over time, $e_{11}$ evolves from being an ``add-on'' reconnaissance vector to part of a targeted attack on the Drupal ecosystem.} 
	\label{fig:time_elapse}
\end{figure*}

This paper presents the first approach to characterize not only single security events, but the context in which they are used in the wild.
The problem of characterizing the evolution of security events, however, is a complex one and presents multiple challenges. 
To illustrate its complexity, consider the real-world example in Figure~\ref{fig:time_elapse}, showing several machines undergoing two coordinated attacks across time, $C_1$ and $C_2$. Both attacks leverage the attack step $e_{11}$, ``CVE-2018-7602 Drupal core RCE.'' $C_1$: \{$e_4$, $e_{10}$, $e_{11}$, $e_{12}$\} mainly functions as a reconnaissance attack including ``Joomla JCE security bypass and XSS vulnerabilities'' ($e_4$), ``Wordpress RevSlider/ShowBiz security byPass'' ($e_{10}$) and ``Symposium plugin shell upload'' ($e_{12}$), together with $e_{11}$. $C_2$: \{$e_7$, $e_5$, $e_{11}$, $e_6$\}, is an attack targeted at the  Drupal ecosystem, consisting of ``phpMyAdmin RFI CVE-2018-12613'' ($e_7$), ``Drupal SQL Injection CVE-2014-3704'' ($e_5$), and ``Apache Flex BlazeDS RCE CVE-2017-3066'' ($e_6$), and the aforementioned $e_{11}$. 
Our goal is to develop a system that allows to automatically analyze the context in which $e_{11}$ is exploited, and identify changing trends.

The first challenge that we can immediately notice from Figure~\ref{fig:time_elapse} is that even though the machines at a certain timestamp are going through the same type of attack (\eg $C_1$ at $t_i$), there are no obvious event relationships reflected in the telemetry recorded by the IPS due to noise (\eg other security events not related to the coordinated attack observed, or certain events relating to the coordinated attack being not observed). If we take the IPS data recorded at timestamp $t_i$, it is not trivial to understand how $e_{11}$ is leveraged by the attackers by directly inspecting the security events, what attack vectors are used together with $e_{11}$, etc. Additionally, it is worth noting that not all security events may be observed in a given observation period. For example, $e_7$ is not observed until timestamp $t_j$.  

The second challenge is that attacks change over time, consequently, the context of a security event and its relationship with other attack steps may drift. It is possible that $C_1$ and $C_2$ can be operated by the same attackers, and that at some point they changed their attack scripts to leverage newly disclosed vulnerabilities (\ie, phpMyAdmin RFI CVE-2018-12613 ($e_7$)). As we can see in Figure~\ref{fig:time_elapse}, from timestamp $t_i$ to $t_j$ attack $C_1$ gradually migrated to or was replaced by attack $C_2$.  However, it is difficult to determine if these new relationships (\eg $e_{11}$ starting to appear in close proximity of $e_5$) at timestamp $t_k$ with respect to those of timestamp $t_i$ are due to noise or are actually indicators of a change in the way $e_{11}$ is being used in the wild. 
Considering all these temporal factors, it is desirable to have a model that is able to understand the context of a security event and its changes over time, and whose output can be quantitatively measured and studied.
This is what \approach aims to do.

\noindent \textbf{Problem formulation.} We formalize our temporal security event evolution approach as follows. A security event $e_{i} \in \mathcal{E}$ is a timestamped observation recorded at timestamp $i$, where $\mathcal{E}$ denotes the set of all unique events and $|\mathcal{E}|$ denotes the size of $\mathcal{E}$. A security event sequence observed in an endpoint $s_j$ is a sequence of events ordered by their observation time, $s_j=\{e^{(j)}_{1}, e^{(j)}_{2}, ..., e^{(j)}_{l}\}$. Let $\mathcal{S}_t=\{s_1^t, ..., s_i^t, ..., s_z^t\}$ denote the set of the security 
events from $z$ endpoints during the $t$-th observation period. Finally we denote $\mathcal{S} = \{ \mathcal{S}_1, ..., \mathcal{S}_t ,..., \mathcal{S}_T \}$, $t=1, ..., T$, as the total security events over time $T$. It is worth noting that not all security events may be observed in a given $\mathcal{S}_t$. 
For example, security events associated with CVEs reported in 2018 are not present in the set of security events collected in 2017. Our goal is to find a mapping function $\mathcal{M}(e_i, \mathcal{S}, T) \rightarrow \{ \eta_{e_i}^t \}$, where $t=1, ..., T$ and $\eta_{e_i}^t \in \mathbb{R}^d$, $d \ll |\mathcal{E}|$ denotes a $d$-dimensional vector representation of the security event $e_i$ at timestamp $t$. 
In the next section, we describe the data used in this paper in more detail.
Then, in Section~\ref{sec:methodology} we describe the methodology used by \approach.

\section{Dataset}
\label{sec:dataset}
\begin{figure}[t]
	\centering
	\includegraphics[width=0.8\linewidth]{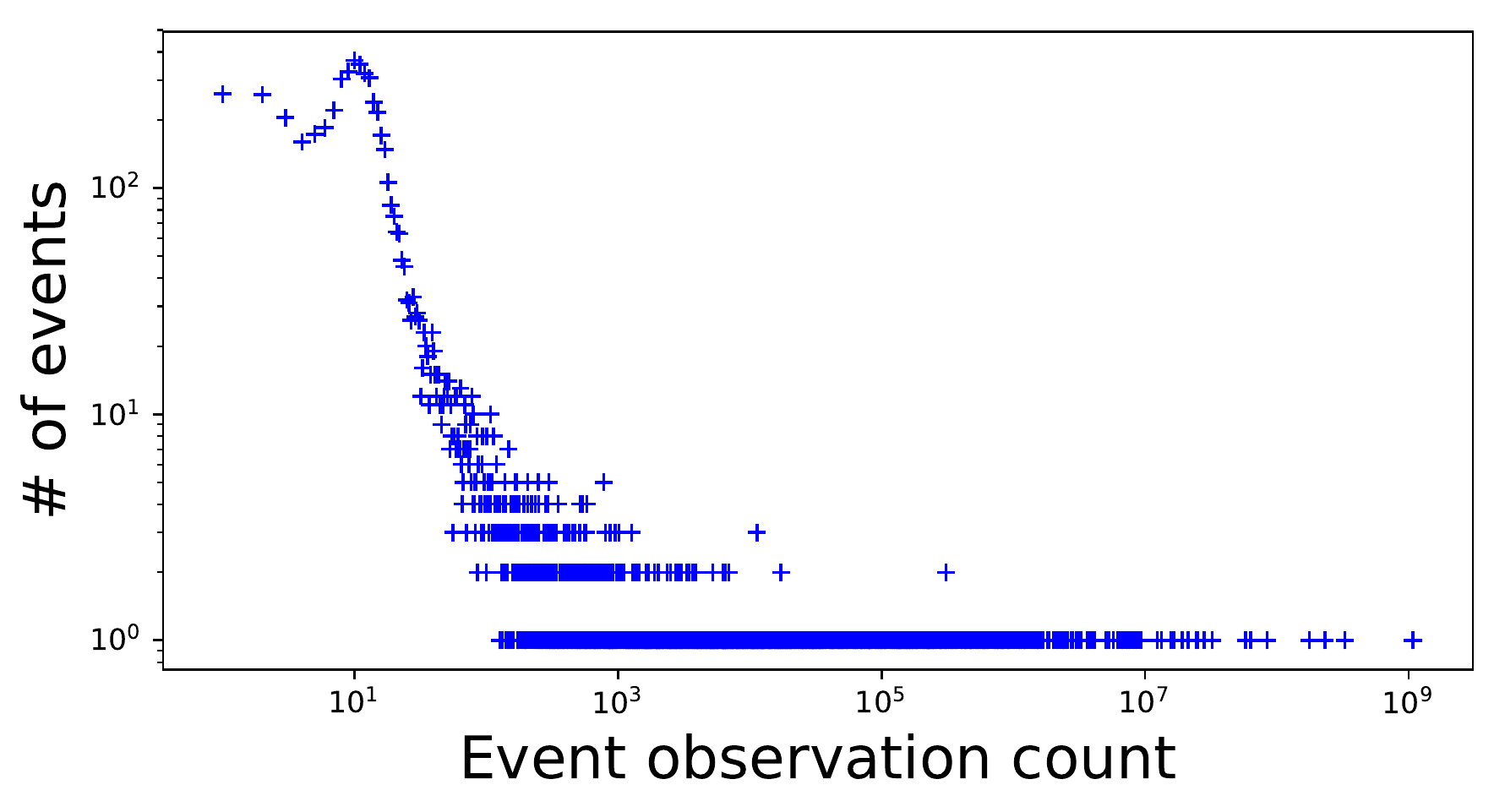}
	\caption{The distribution of events in our IPS security event dataset follows a power-law, much like the distribution of words in natural language, confirming the appropriateness of word embeddings to study the evolution of the use of security events over time.}
	\label{fig:powerlaw}
\end{figure}

\noindent \textbf{Data origin.} As a proxy for the attack steps performed by miscreants in the wild, we use security event data collected from Symantec's intrusion prevention system (IPS). The company offers end users to explicitly opt in to its data sharing program to help improving its detection capabilities. To preserve the anonymity of users, endpoint identifiers are anonymized and it is not possible to link the collected data back to the users that originated it. Meta-information associated with a security event is recorded when the product detects network-level or system-level activity that matches a predefined signature (\ie a security event). 

\noindent \textbf{Data collection.} To thoroughly investigate security event evolution, we collected 102 days (one observation day per week for 102 consecutive weeks) of data between December 1, 2016 and November 08, 2018. From this data we extract the following information: anonymized machine ID, timestamp, security event ID, event description, system actions, etc. On average, we collect 190 million security events collected from tens of millions unique machines per day. These security events were then reconstructed on a per machine basis and sorted chronologically. Note that for privacy reasons we use the anonymized endpoint ID to reconstruct a series of security events detected in a given mahcine and discard it after the reconstruction process is done. In total, the monitored machines generated 8,087 unique security events over the 102 observation days.

\noindent\textbf{Data Limitations.} It is important to note that the security event data is collected passively. That is, these security events are recorded only when corresponding attack signatures are triggered. Any events preemptively blocked by other security products cannot be observed. Additionally, any events that did not match the predefined signatures are also not observed. Hence the findings in this paper reflect security event evolution observed by Symantec's IPS, and the data can only be considered as a proxy for the actual attacker behavior in the wild.
For example, we are unable to trace how zero day attacks are exploited in the wild~\cite{bilge2012before}. 
However, as we show in Section~\ref{sec:quan_eval} this data still allows \approach to identify meaningful trends in how security events are used and evolve.
Additionally, \approach could be applied to any dataset with similar characteristics (\ie a sequence of security events).
Another limitation is that our dataset is composed of weekly snapshots, and we are therefore unable to characterize the evolution of security events that are faster than that.
While this could prevent us from detecting quick anomalies in the way security events are used (\ie those that go back to ``normal'' in a matter of a few days), this data is still representative enough to identify long term trends.
We provide a more detailed discussion on the limitations of our data in Section~\ref{sec:discussion}. 

\noindent\textbf{Appropriateness for word embeddings.}
As we mentioned, the word embedding techniques used by \approach come from the natural language processing field.
Word frequency in natural language follows a power-law distribution, and techniques from language modeling account for this distributional behavior. 
For these techniques to be appropriate to our data, therefore, it is ideal that our security events  follow a similar distribution.
Figure~\ref{fig:powerlaw} shows that the events in our dataset indeed follows a power-law distribution.
This similarity forms a solid theoretical foundation for us to use word embedding techniques to encode latent forms of security events, by considering sequences of security events in the IPS logs as short sentences and phrases in a special language. 
In the next section, we describe how \approach builds temporal embeddings from a sequence of security events in detail. 

\section{Methodology}
\label{sec:methodology}
\begin{figure}[t]
	\centering
	\includegraphics[width=\linewidth]{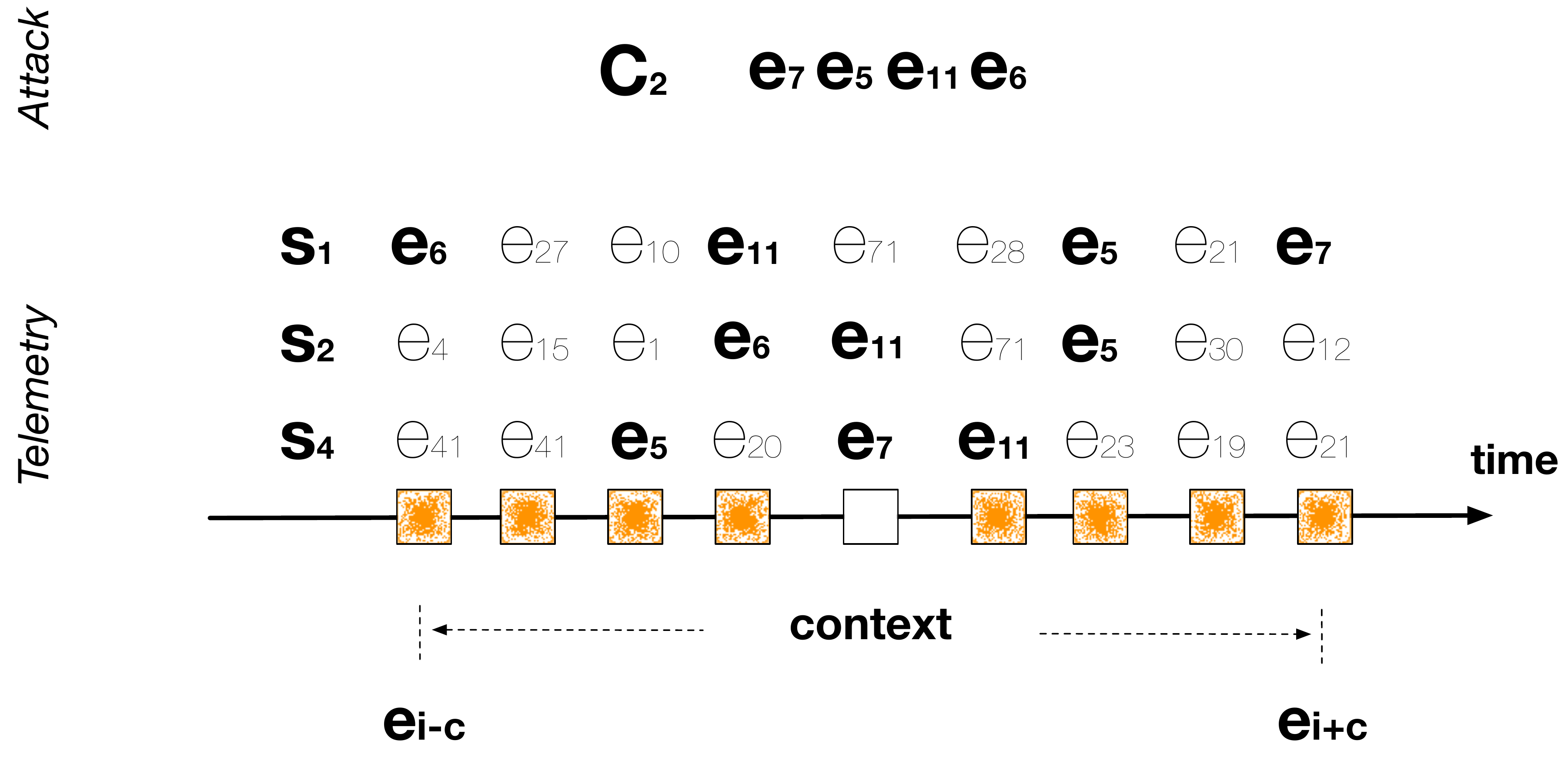}
	\caption{Illustration of context window ($c=4$). }
	\label{fig:context_window}
\end{figure}

In this section we first define the context window used in this work.
We then formalize the techniques used to generate vector embeddings of security events.
Finally, we describe \approach's architecture.

\subsection{Context Window} 
\label{sec:context_window}

Previous research made several interesting observations that different attack vectors are often packed together by attackers for a given period of time. For example, Kwon \etal~\cite{kwon2017catching} observed that silent delivery campaigns exhibit synchronized activity among a group of downloaders or domains and access the same set of domains to retrieve payloads within a short bounded time period. Shen \etal~\cite{shen2018tiresias} pointed out that some machines may potentially observe different attacks from various adversary groups happening at the same time, and one coordinated attack may be observed by different machines. On the defense side (\ie IPS telemetry), we consequently observe that related security events co-occur within a context (\ie the sequence of attack steps that are commonly performed together with an attack step of interest). Note that this context can be defined as a time window~\cite{kwon2017catching} or a rollback window~\cite{shen2018tiresias}. 

In this paper, we define the context as a sliding window, denoted as $c$, centering around a given security event $e_i$ (see Figure~\ref{fig:context_window}). The purpose of using this symmetric context window is to deal with the noise incurred by concurrency at the telemetry level (see Section~\ref{sec:motivation}). For example, given a real-world coordinated attack $e_7,e_{5},e_{11},e_{6}$ (highlighted in bold in Figure~\ref{fig:context_window}), each endpoint may observe the attack vectors in different order (\eg $e_7$ and $e_{5}$ may switch orders), attack vectors might be diluted by other unrelated security events (\eg $e_{71}$ observed between $e_6$ and $e_{5}$ in $s_2$), or certain security events are not observed, for example because they have been blocked by other security products before the IPS was able to log them (\eg $e_6$ not observed in $s_4$). The proposed context window mechanism is able to capture the  events surrounding a given security event (\ie before and after), minimizing the impact of noise incurred by concurrency.

\subsection{Temporal Security Event Embedding} 
\label{sec:temporal_embedding_method}

The proposed temporal security event embedding  is adapted from dynamic word embeddings by Yao \etal~\cite{yao2018dynamic}. We use pointwise mutual information (PMI), a popular measure for word associations, to calculate weights between two security events given a contextual window $c$ and an observation period $t$. PMI measures the extent to which the events co-occur more than by chance or are independent. The assumption is that if two events co-occur more than expected under independence there must be some kind of relationship between them. For each $t$-th observation period,  we build a $|\mathcal{E}| \times |\mathcal{E}|$ PMI matrix, where a PMI value between $e_i$ and $e_j$ is defined as follows.

\begin{align}
PMI_t(e_i, e_j, c, \mathcal{S}) &= max (log \left( \frac{p_t(e_i, e_j)}{p_t(e_i)p(e_j)} \right), 0), \nonumber \\
p_t(e_i, e_j)     &= \frac{W(e_i, e_j)}{|\mathcal{S}_t|}, \nonumber \\
p_t(e_i)          &= \frac{W(e_i) }{|\mathcal{S}_t|},  \label{eq:PMI}
\end{align}

\noindent  where $W(e_i)$ and $W(e_j)$ respectively count the occurrences of security events $e_i$ and $e_j$ in $\mathcal{S}_t$, and $W(e_i, e_j)$ counts the number of times $e_i$ and $e_j$ co-occur within a context window (see Figure~\ref{fig:context_window}, Section~\ref{sec:context_window}) in $\mathcal{S}_t$. Note that when $W(e_i, e_j)$, the number of times $e_i$ and $e_j$ co-occurring in a given contextual window is small, $log \left( \frac{p_t(e_i, e_j)}{p_t(e_i)p(e_j)} \right)$ can be negative and affects the numerical stability. Therefore, we only keep the positive values in Eq~\ref{eq:PMI} (see~\cite{levy2014neural}). 

Following the definition of $PMI_t$, the security event embedding $H(t)$, e.g., $\eta_{e_i}^t \in H(t)$, at $t$-th observation time is defined as a factorization of $PMI_t(c, \mathcal{S})$,

\begin{equation}
	H(t) H(t)^T \approx PMI_t(c, \mathcal{S}).
\end{equation}

\noindent The denser representation $H(t)$ reduces the noise~\cite{rapp2003word} and is able to capture events with high-order co-occurrence (\ie that appear in similar contexts)~\cite{mikolov2013distributed, pennington2014glove}. These characteristics enable us to use word embedding techniques to encode latent forms of security events, and interpret the security event evolution in a meaningful and measurable way. Note that Li~\etal~\cite{li2015word} and Levy~\etal~\cite{levy2014neural} have theoretically proven that the skip-gram negative sampling (SGNS) used by the word2vec model can be viewed as explicitly (implicitly) factorizing a word co-occurrence matrix. We refer interested readers to ~\cite{li2015word,levy2014neural} for theoretical proofs.

Across time $T$, we also require that $\eta_{e_i}^t \approx \eta_{e_i}^{t+1}$. This means that the same security event should be placed in the same latent space so that their changes across time can be reliably studied. This requirement roots upon a practical implication. For example, a security event was observed after its associated CVE was disclosed. Its embeddings must therefore approximately stay the same before the disclosure date. Otherwise, we would observe unwanted embedding changes and invalidate the findings. To this end, Yao \etal~\cite{yao2018dynamic} identified the solution of the following joint optimization problem as the temporal embedding results. Note that throughout this section, $\Vert . \Vert$ denotes squared Frobenius norm of a vector. 

\begin{equation}
\label{eq:dwe}
\begin{split}
 \min_{H(1),...,H(T)} \frac{1}{2} & \sum_{t=1}^{T}  \Vert PMI_t(c, \mathcal{S}) - H(t) H(t)^T \Vert_2  \\
       & + \frac{\alpha}{2} \sum_{t=1}^{T}  \Vert H(t)  \Vert^2 + \frac{\beta}{2}  \sum_{t=1}^{T} \Vert H(t-1) - H(t)  \Vert^2,
\end{split}
\end{equation}

\noindent where $\alpha$ and $\beta$ are parameters respectively regularizing $H(t)$, and making sure that $H(t-1)$ and $H(t)$ are aligned (\ie embeddings should be close if their associated contexts don't change between subsequent times.). In this way, all embeddings across time $T$ are taken into consideration. At the same time, this method can accommodate extreme cases such as the one in which security event $e_i$ is not observed in $\mathcal(S)_t$ since the optimization is applied across all time slices in Eq~\ref{eq:dwe}. We refer interested readers to~\cite{yao2018dynamic} for theoretical proofs and empirical comparison studies with other state-of-the-art embedding approaches. Following~\cite{yao2018dynamic}, we use grid search to identify the best parameters and experimentally set $\alpha=10$, $\beta=40$, $c=8$, $d=50$ and run 5 epochs for all the evaluations throughout our paper. 

\subsection{\approach Architecture} 
\label{sec:architecture}

\begin{figure*}[ht]
	\centering
	\includegraphics[width=\linewidth]{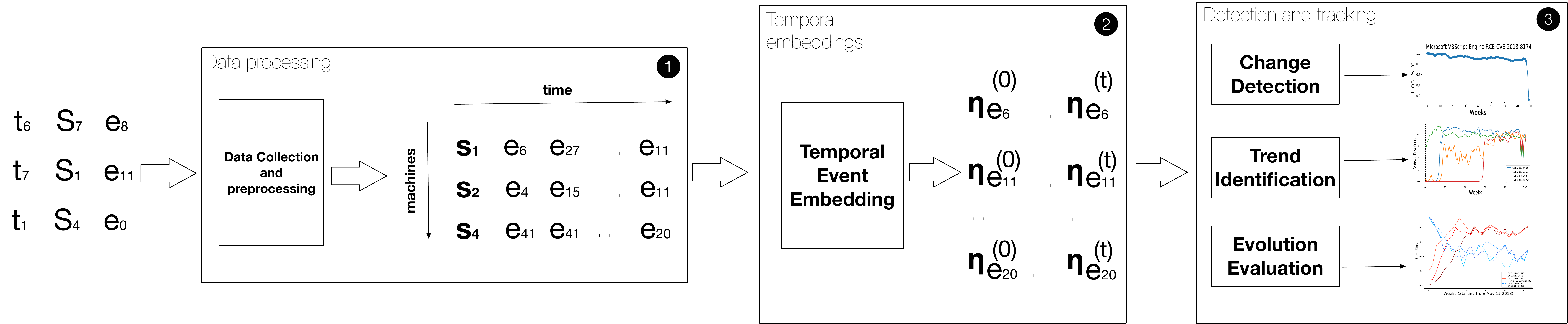}
	\caption{\approach's Architecture.}
	\label{fig:architecture}
\end{figure*}

The architecture and workflow of \approach is depicted in Figure~\ref{fig:architecture}. Its operation consists of three phases: \ding{182} data collection and preprocessing, \ding{183} temporal event embedding, and \ding{184} event tracking and monitoring.

\noindent\textbf{Data collection and preprocessing (\ding{182}).} \approach takes the security event stream generated by endpoints (\eg computers that installed an IPS). The goal of the data collection and preprocessing module is to prepare the data for the temporal event embedding method detailed in Section~\ref{sec:temporal_embedding_method}. \approach then consumes this timestamped security event data generated from millions of machines that send back their activity reports. The collection and preprocessing module reconstructs the security events observed on a given machine $s_j$ as a sequence of events ordered by timestamps, in the format of $s_j=\{e^{(j)}_{1}, e^{(j)}_{2}, ..., e^{(j)}_{l}\}$. The output of the data collection and preprocessing module is $\mathcal{S}_t=\{s_1^t, ..., s_i^t, ..., s_z^t\}$ where $z$ denotes the number of machines.

\noindent\textbf{Temporal event embedding (\ding{183}).} The core operation of \approach is embedding these security events into a low dimensional space over time. This phase takes $\mathcal{S}$ as input and encodes latent forms of security events, by considering sequences of security events in the IPS logs as short sentences and phrases in a special language. In this way, each security event, at a timestamp $t$, is represented by a $d$-dimensional vector representation $\eta_{e_i}^t$, and later aligned across time.

\noindent\textbf{Detection and monitoring (\ding{184}).} Once the security events are encoded in low-dimensional space, \approach is able to use various metrics (Section~\ref{sec:metric}) to detect changes (Section~\ref{sec:fidelity}), identify event trends (Section~\ref{sec:robustness}), and monitor how security events are exploited in the wild (Section~\ref{sec:case1}) in a measurable and quantifiable way. 

\section{Evaluation}
\label{sec:quan_eval}

In this section, we provide a thorough evaluation of temporal event embeddings and \approach. We designed a number of experiments that allow us to answer the following questions:

\begin{itemize}

  \item Can we use the temporal embeddings calculated by \approach to identify changes in how a security event is used in the wild (see Section~\ref{sec:fidelity})? 
    To this end, we need our temporal embeddings to present high fidelity over time.
    The rationale behind this question is that the same security event should be placed in the same latent space by the proposed temporal event embedding method (see Section~\ref{sec:methodology}). Their changes across time can be reliably studied (see Section~\ref{sec:evolution}). 
	
  \item Can we leverage temporal embeddings to identify trends in the use of security events (see Section~\ref{sec:robustness})? 
    The rationale behind this evaluation is that embedding vector norms across time should be more robust to the changes than word frequency which is static (\ie calculated at a specific point of time) and sporadic. 
	
  \item Can we leverage temporal embeddings to meaningfully understand the evolution of security events, and monitor how security events are exploited in the wild (see Section~\ref{sec:case1})? 

\end{itemize}

In the following, we first define the metrics used by our evaluation.
We then proceed to show that \mbox{\approach} is effective in answering these three research questions, and discuss the performance of our approach, showing that \mbox{\approach} is able to process a day of data within minutes. Finally, we present further evaluation of \approach, showing an end-to-end case on how our system can be used to assess the evolution in the use of a specific vulnerability in the wild.

\subsection{Evaluation Metric}
\label{sec:metric}
We use \emph{cosine similarity} as the distance metric to quantify the temporal embedding changes at time $t$ in the latent space. That is, for any two embeddings (\ie $\eta_{e_i}^{(t)}$ and $\eta_{e_j}^{(t)}$), the similarity is measured as

\begin{equation} \label{equ:cosine}
similarity(\eta_{e_i}^{(t)}, \eta_{e_j}^{(t)})= \frac{{\eta_{e_i}^{(t)}}^T \eta_{e_j}^{(t)}}{ \Vert \eta_{e_i}^{(t)} \Vert_2 \Vert \eta_{e_j}^{(t)} \Vert_2 }.
\end{equation}

\noindent Note that in this paper the cosine similarity is used in positive space, where the outcome is bounded in [0, 1]. That is, two vectors with the same orientation have a cosine similarity of 1 (most similar), two vectors oriented at 90\degree~relative to each other have a similarity of 0 (not similar).

Following Eq~\ref{equ:cosine}, we denote the \emph{neighborhood} of a security event embedding $e_i^{(t)}$ as $\mathcal{N}(e_i^{(t)})$, and accordingly defined as

\begin{equation} \label{equ:neighbor}
	\mathcal{N}(e_i^{(t)}) = argsort_{e_j^{(t)}} (similarity(e_i^{(t)}, e_j^{(t)})).
\end{equation}

\noindent $\mathcal{N}(e_i^{(t)})$ enables us to use temporal embeddings to discover and analyze how different security events are used together with $e_i$. 
We use $\mathcal{N}_k(e_i^{(t)})$ to denote the top $k$ closest neighbors of $e_i$.
As we show in Section~\ref{sec:case1}, this can be used to identify security events that are frequently used together as part of a multi-step attack.

We also use a \emph{weighted drift} metric to measure a security event \emph{relative} changes.
This metric is defined in Eq~\ref{eq:changepoint} as

\begin{equation}
\label{eq:changepoint}
weighted\_drift(e_i) = argsort_t \left( \frac{ \Vert \eta_{e_i}^{(t-1)}, \eta_{e_i}^{(t)} \Vert}{ \sum_{e \in \mathcal{E}} \Vert \eta_{e}^{(t-1)}, \eta_{e}^{(t)} \Vert }   \right).
\end{equation}

\noindent Eq~\ref{eq:changepoint} normalizes a security event's embedding change by the sum of all security event changes within that observation period. This metric enables us to measure how a security event changes comparing to the other security events within a given observation point.

\subsection{Change Detection}
\label{sec:fidelity}
\begin{figure}[t]

    \begin{subfigure}{0.92\linewidth}
        \centering
        \includegraphics[width=\linewidth]{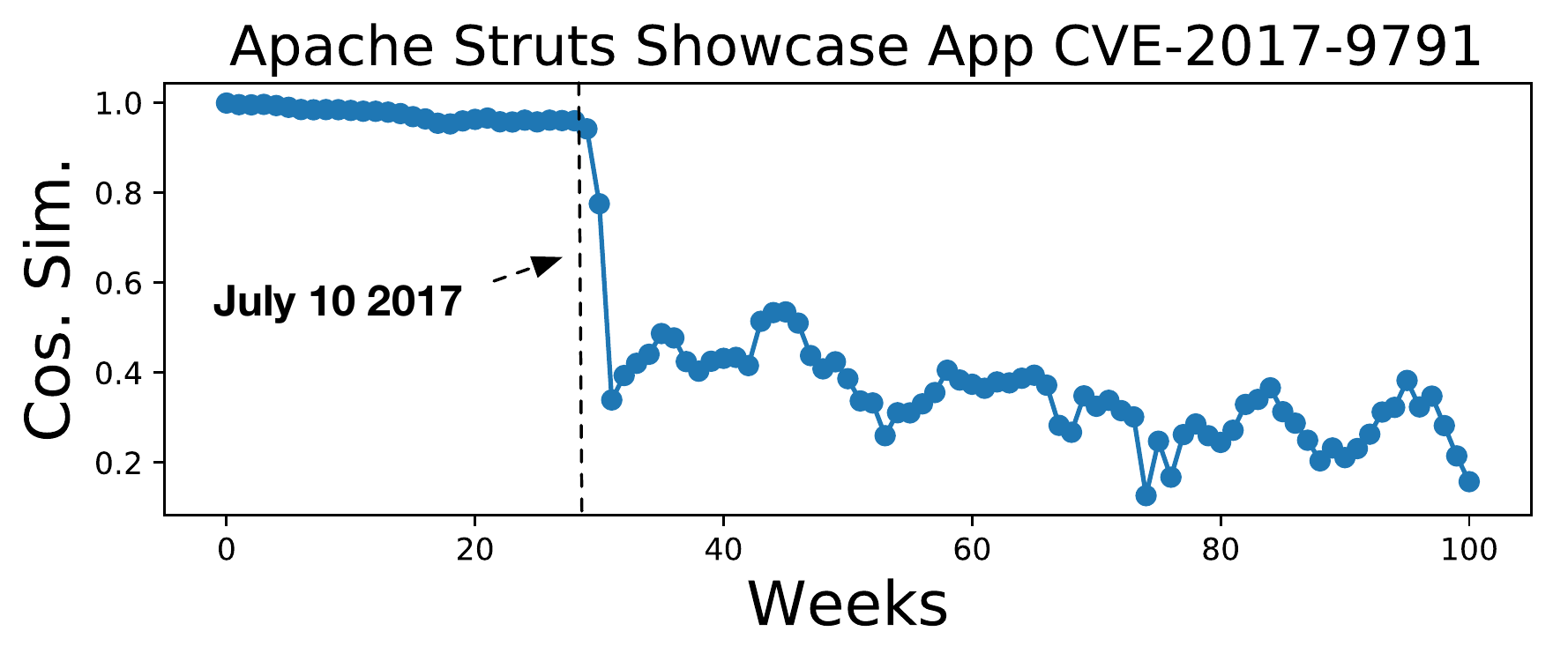}
        \caption{CVE-2017-9791 disclosure date: July 10 2017}
        \label{fig:30102}
    \end{subfigure}
    \hfill
    \begin{subfigure}{0.92\linewidth}
        \centering
        \includegraphics[width=\linewidth]{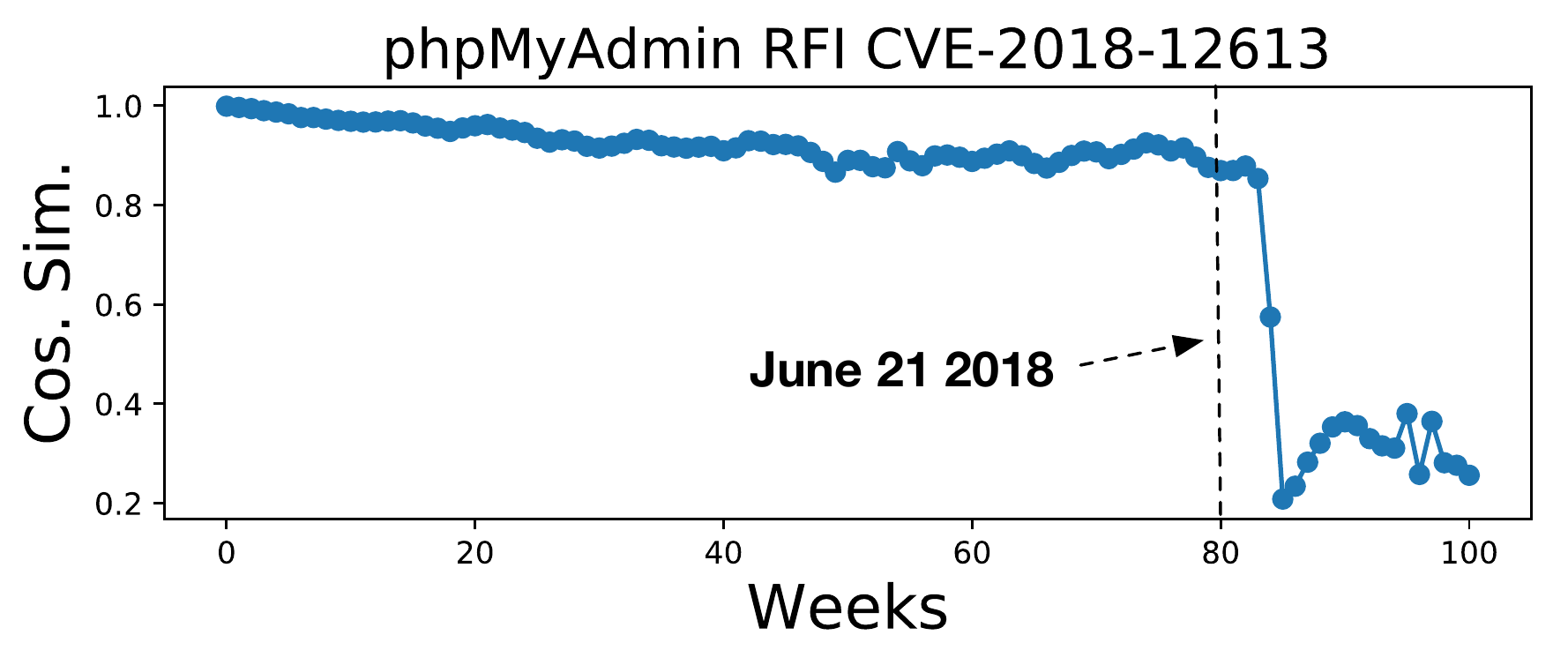}
        \caption{CVE-2018-12613 disclosure date: June 21 2018}
        \label{fig:30910}
    \end{subfigure}
	\hfill
    \begin{subfigure}{0.92\linewidth}
        \centering
        \includegraphics[width=\linewidth]{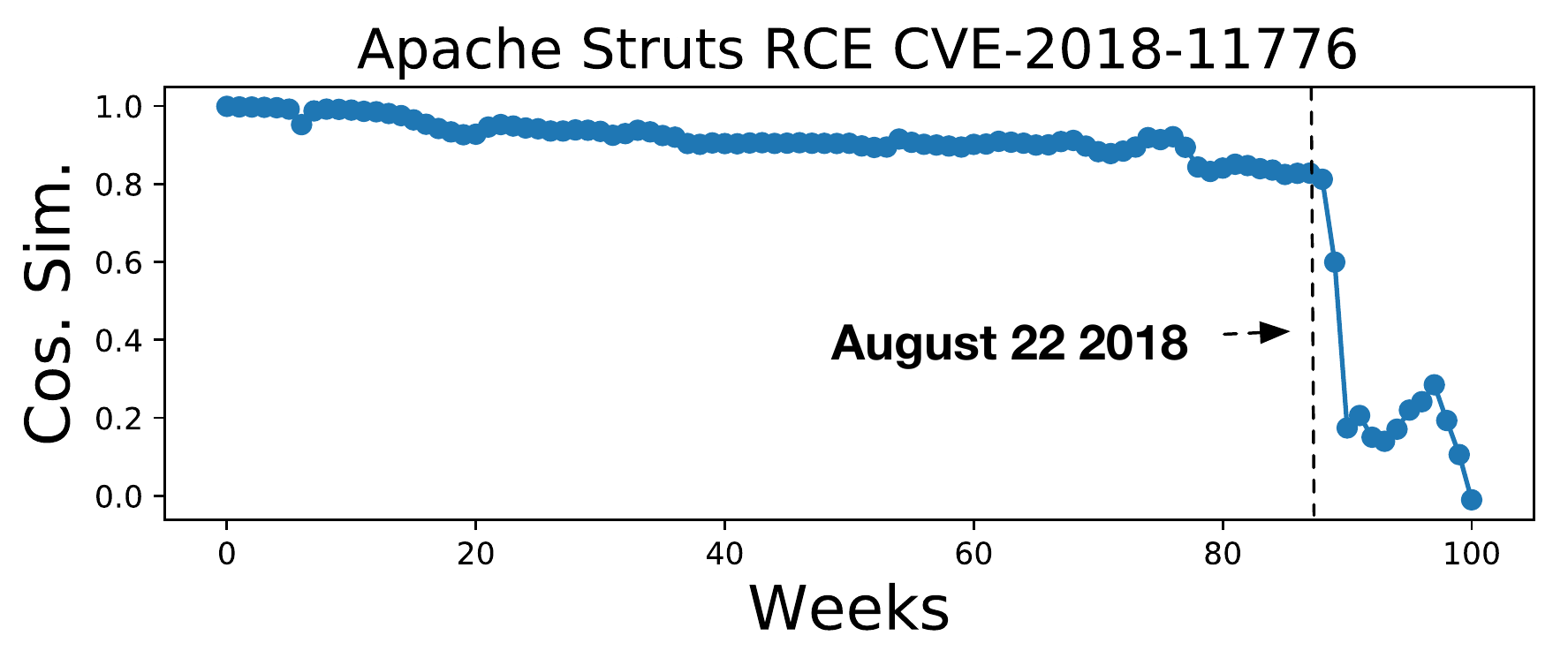}
        \caption{CVE-2018-11776 disclosure date: August 22 2018}
        \label{fig:30962}
    \end{subfigure}
	\caption{Temporal embedding results of ``Apache Struts Showcase App CVE-2017-9791'' (\ref{fig:30102}), ``phpMyAdmin RFI CVE-2018-12613'' (\ref{fig:30910}), and ``Apache Struts RCE CVE-2018-11776'' (\ref{fig:30962}). 
	The cosine similarities of the CVE embeddings are stable before they are publicly disclosed, and decline swiftly after the disclosure.}
	\label{fig:fidelity} 
\end{figure}

One of the key practical questions when evaluating the temporal security event embeddings built by \approach is determining the fidelity of the embedding results over time. 
In this paper, fidelity refers to the condition that the same security event should be placed in the same latent space. That is, if the frequency and the contexts of a security event between subsequent time slices don't change, its latent embedding should stay the same. This consistency allows the change to be reliably detected.
This requirement lays the foundation to quantitatively study their changes. The concept of fidelity is different from the stability term used in previous research approaches in which stability was used to evaluate how classifiers perform after certain period of time.  Bearing this difference in mind, we use the following two criteria to evaluate the fidelity of temporal embeddings and show how \approach can faithfully capture both single event usage change and global changes: 

\begin{itemize}
	
	\item \textbf{criterion a.} The cosine similarity of the event embeddings must be stable when an event usage does not change between subsequent time slices. 
	
	\item \textbf{criterion b.} The cosine similarity of these embeddings should change swiftly if these events are used in different attacks or emerge as a new attack vector. 
	
\end{itemize}

It is important to note that while these criteria are helpful in demonstrating the power of word embeddings extracted by \approach, they are self-referential and not in themselves sufficient to validate the effectiveness of our approach.

\noindent \textbf{Single event change detection.} To evaluate whether our two criteria hold for our dataset, we use three CVEs, ``Apache Struts Showcase App CVE-2017-9791'' (Figure~\ref{fig:30102}), ``phpMyAdmin RFI CVE-2018-12613'' (Figure~\ref{fig:30910}), and ``Apache Struts RCE CVE-2018-11776'' (Figure~\ref{fig:30962}). 
These CVEs were disclosed between 2017 and 2018.  
Regarding the aforementioned two evaluation criteria, these vulnerabilities were not disclosed in 2016, and therefore they did not have a matching signature in the IPS from which we collected our data.
Thus, they form a good baseline for temporal fidelity evaluation. We therefore expect the following properties to hold:

\begin{description}
	
	\item[response to a.] Before a vulnerability was disclosed, its corresponding signature does not exist hence its non-existent context should stay the same until timestamp $t$. 
	  That is, if the vulnerability's disclosure date is $t$, $similarity(\eta_{e_i}^{(0)}, \eta_{e_i}^{(z)})$, where $z \in (0, t]$, should be stable. 
	
	\item[response to b.] After the disclosure date, the cosine similarity values of its embeddings should change swiftly. The justification is obvious.  If attackers start exploiting a vulnerability, its corresponding security event moves away from its non-existent context and such drift leads to embedding changes.  
	
\end{description}

\noindent For each CVE, we calculate the cosine similarity between each event’s current representation (\ie at timestamp $t$) and its \emph{original} representation (\ie at timestamp 0, on December 1 2016) over our observation period (\ie $similarity(\eta_{e_i}^{(0)}, \eta_{e_i}^{(t)})$, where $t=1...T$. See Eq~\ref{equ:cosine}). The results are shown in Figure~\ref{fig:fidelity}. As we can observe, the temporal embeddings of CVE-2017-9791, CVE-2018-12613 and CVE-2018-11776 are stable across time and their cosine similarity values are above 0.9 before their respective disclosure dates (see \textbf{criterion a}). 
The way to interpret \textbf{criterion a} is that before a vulnerability is disclosed, its corresponding signature does not exist, and therefore its context is \emph{non-existing}. 
As such, this context should remain constant until the vulnerability starts being exploited in the wild.
Figure~\ref{fig:30102} shows that the cosine similarity between the embeddings of CVE-2017-9791 calculated daily and the original one recorded on day 0 is stable and above 0.95 before July 10 2017, which is when the vulnerability was disclosed. 
Note that the similarity is not strictly 1.0 because of marginal deviation incurred by joint optimization across time slices (see Eq~\ref{eq:dwe}). 
Nevertheless, the high similarity before the disclosure date shows that \mbox{\approach} obtains correct temporal embeddings. 
After their public disclosure of each CVE, on the other hand, we expect the context in which each vulnerability is exploited to quickly change. 
This can be measured by \mbox{\approach} with the fact that the cosine similarity values of CVE-2017-9791, CVE-2018-12613, and CVE-2018-11776  decline quickly and move away from the original non-existing context built for those CVEs (see \textbf{criterion b}). 
This phenomenon exemplifies that the temporal embeddings capture the changes in the context in which a security event is used.

It is also worth noting that the temporal embeddings of CVE-2018-11776 show an immediate change after disclosure, while those of CVE-2018-12613 are slightly delayed for a couple of weeks (\ie CVE-2018-12613 was officially published on June 21 2018 and the embedding starts to drift on July 12 2018). This phenomenon, \ie the gap between public disclosure dates and real world exploits was well discussed in Sabottke \etal~\cite{sabottke2015vulnerability}, and \approach allows to easily observe it. 

\begin{figure}[t]
	\centering
	\includegraphics[width=\linewidth]{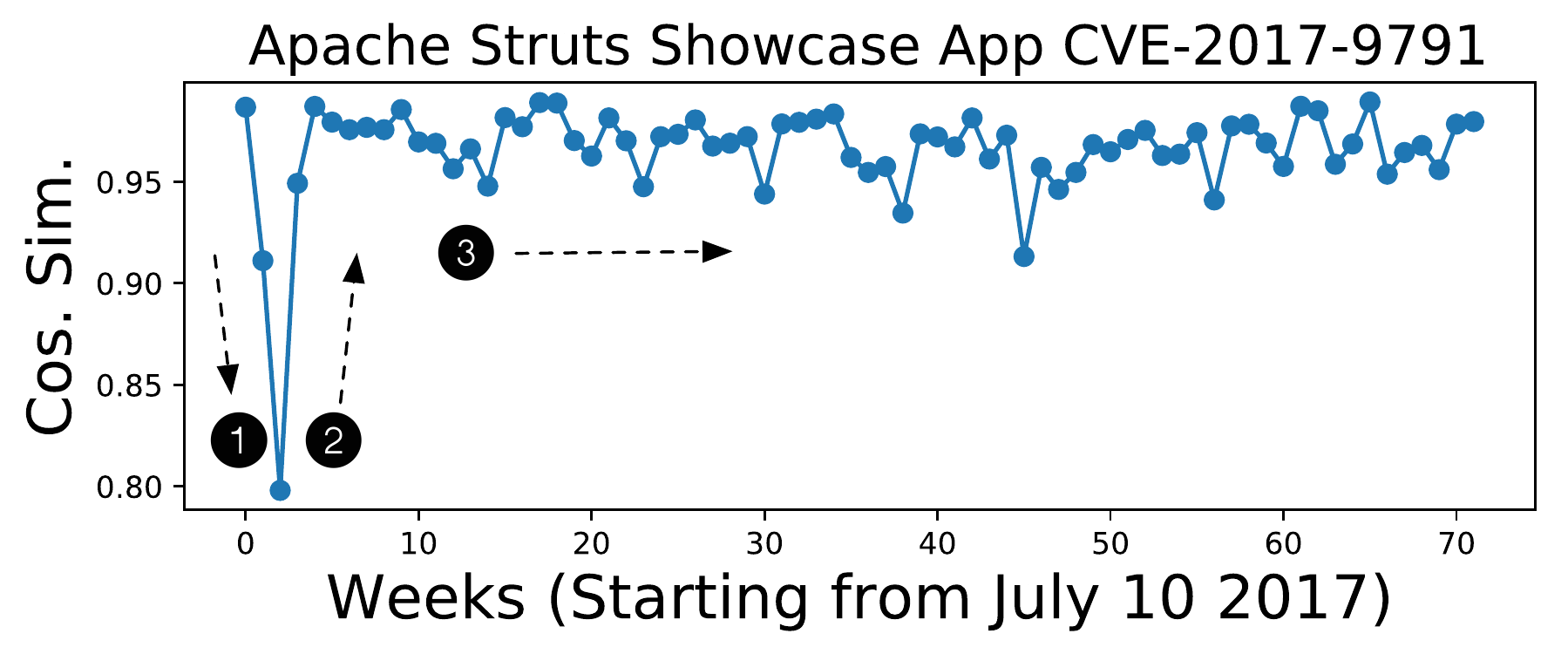}
	\caption{Temporal embedding result of Apache Struts Showcase App CVE-2017-9791. Cosine similarity values for each plot is calculated as $similarity(\eta_{e_i}^{(t-1)}, \eta_{e_i}^{(t)})$ (\ie cosine similarity between subsequent time slices), where $t$ starts from July 10 2017 (the public disclosure date). }
	\label{fig:30102_dynamic}
\end{figure}

The temporal embeddings generated by \approach not only allow us to identify when a vulnerability starts being exploited in the wild, but also how 
these event embeddings change after the disclosure date.
To monitor and evaluate these changes, instead of comparing the context of a security event with the one extracted from the first day of observation, we compare the cosine similarity of the contexts extracted on subsequent time slices -- between each event’s current representation (\ie at timestamp $t$) and its previous representation (\ie at timestamp $t-1$). In short, we calculate $similarity(\eta_{e_i}^{(t)}, \eta_{e_i}^{(t-1)})$ (see Eq~\ref{equ:cosine}), which enables us to capture how the context of each event evolves between two subsequent observations. If the use of an event remains stable, the cosine similarity between $\eta_{e_i}^{(t)}$ and $\eta_{e_i}^{(t-1)}$ will remain high. If, on the other hand, the event experiences a sudden change in the way it is used in the wild, then its context will also significantly change and the cosine similarity with the previous observation will suddenly decrease, allowing an analyst to identify the point in time in which this change happened.
To demonstrate this, we reuse the security event ``Apache Struts Showcase App CVE-2017-9791'' from earlier in this section. We calculate the cosine similarity between subsequent snapshots, where $t_0$ starts from July 10 2017 (the public disclosure date). This evolution is depicted in Figure~\ref{fig:30102_dynamic}.

We can observe the following: \ding{182} The cosine similarity values decline for the first three weeks. This phenomenon implies that CVE-2017-9791 started being exploited in different attacks after its disclosure date, with attackers trying different strategies to reliably exploit this vulnerability. \ding{183} the cosine similarity increases between the 3$rd$ week and the 10$th$ week. This phenomenon implies that CVE-2017-9791 became being exploited in less diversified attacks, indicating that attackers were converging towards a stable way to exploit the CVE. \ding{184} The cosine similarity stabilizes after the 10$th$ week, which means that CVE-2017-9791 started being exploited in a \emph{stable} context. 
This could indicate that attackers weaponized the CVE into a reliable attack, and possibly developed methods to exploit it at scale (e.g., by including it in an exploit kit). 
Later in the timeline, we can see other changes in the way in which attacks are exploited, but after each sudden change we observe a stabilization in how the CVE is exploited, indicating that attackers keep the same modus operandi over long periods of time. 

A possible concern is that the changes in context identified by \mbox{\approach} might be due to noise and not representative of actual changes in the modus operandi of attackers.
To demonstrate that this is not the case, we use the event co-occurrence matrix $PMI_t(c, \mathcal{S})$ as defined in Section~\ref{sec:methodology}. 
This matrix captures the co-occurrence of any two events within the context window. For each observation time $t$, we select the top events that co-occurred with CVE-2017-9791 to better understand the phenomenon. 
If the changes in the use of a CVE identified by \mbox{\approach} are meaningful, we expect the co-occurrence matrix on that day to suddenly change, but to later stabilize and remain similar over time.
For the first three weeks after disclosure in Figure~\ref{fig:30102_dynamic} (\mbox{\ding{182}}), CVE-2017-9791 was used in conjunction with known attack vectors such as Apache Struts RCE CVE-2013-2251, HTTP Apache Tomcat UTF-8 dir traversal CVE-2008-2938, and malicious OGNL expression upload. 
By the third week, while some attack vectors were still associated with CVE-2017-9791, the vulnerability gradually started being used together with more recent server attack vectors (\eg Apache Struts RCE CVE-2016-3087) and application vulnerabilities (\eg WebNMS RCE CVE-2016-6603 and Web CMS Think PHP RCE). 
After \mbox{\ding{183}}, CVE-2017-9791 started being used consistently with the aforementioned attack vectors and with several additional attack vectors (\eg  Apache Struts dynamic method invocation RCE CVE-2016-3081, Drupal PHP RCE, and generic PHP REC) Once CVE-2017-9791 reached \mbox{\ding{184}}, its usage patterns became reasonably stable. 
Note that small fluctuations still happen when new Apache Struts related vectors were disclosed and exploited (\eg. Apache Struts CVE-2017-9805 (week 11), CVE-2017-12611 (week 15) and CVE-2017-12617 (week 21), and temporary withdrawn of CVE-2017-12617 (around week 45) in the attacks. 
These changes are reliably detected by \approach.

In summary, our method is able to capture changes in the security event embeddings with high fidelity.

\begin{figure}[t]
	\centering
	\includegraphics[width=\linewidth]{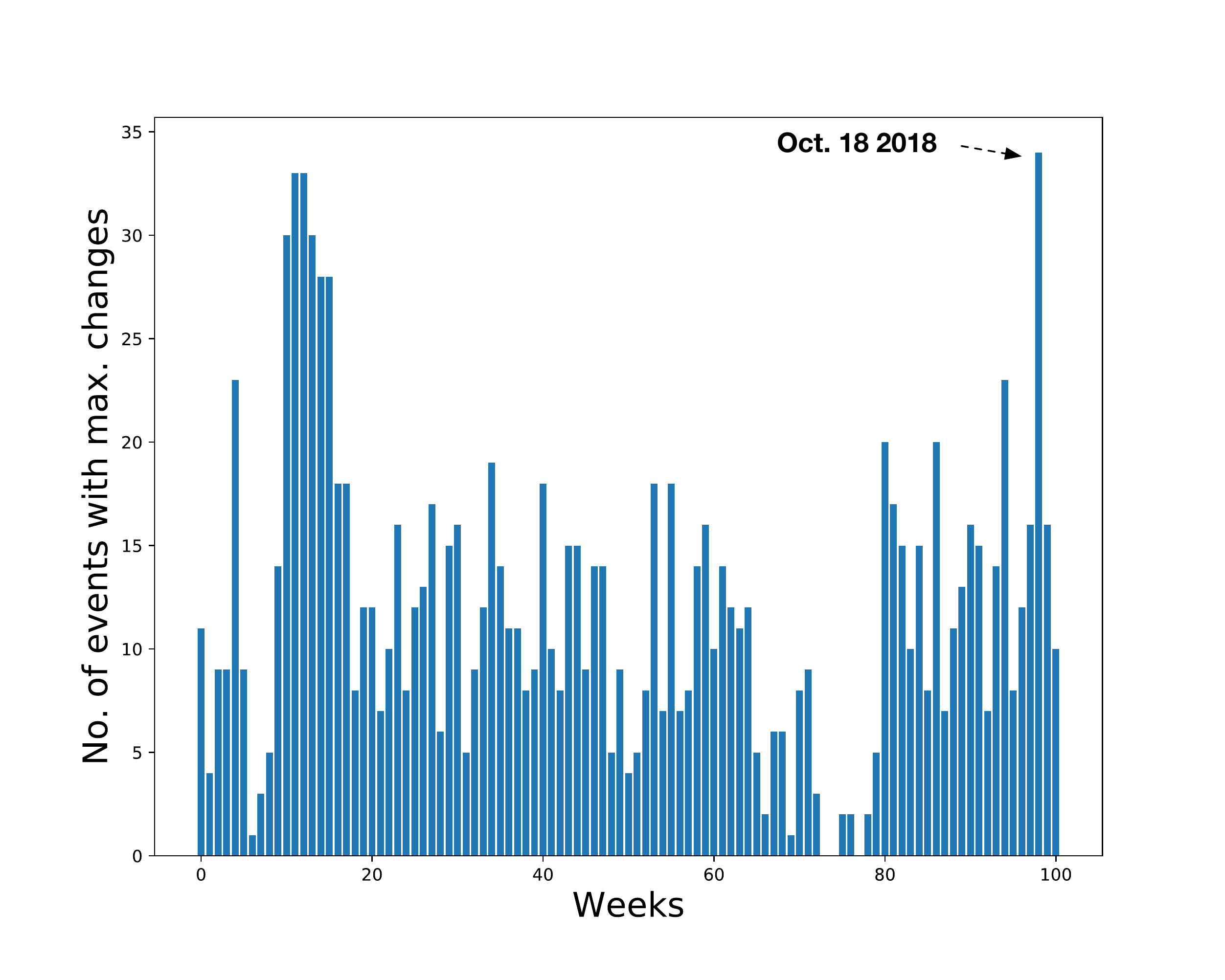}
	\caption{Summary of the security event changes between December 1 2016 and November 15 2018. }
	\label{fig:changepoints}
\end{figure}

\begin{table*}
\centering
\resizebox{0.8\linewidth}{!} {
\begin{tabular}{|l|l|l|l|}
\hline
\begin{tabular}[c]{@{}l@{}}Rank\\ (by changes)\end{tabular} & Mar. 9 2017                                                                                & Jan. 4 2018                  & Oct. 15 2018                                 \\ \hline
1                                                           & ZyNOS Information Disclosure                                                               & Rig Exploit Kit Website      & Unwanted Extension or Scam Sites Redirection \\ \hline
2                                                           & \begin{tabular}[c]{@{}l@{}}WordPress Mobile-Detector \\ Arbitrary File Upload\end{tabular} & Malicious Javascript Website & Fake Tech Support Website                    \\ \hline
3                                                           & \begin{tabular}[c]{@{}l@{}}Netgear Router Remote Command \\ Execute\end{tabular}           & Fake Tech Support Website    & Drupal Core RCE              \\ \hline
4                                                           & Wordpress Arbitrary File Download                                                          & Malicious Redirection        & Fake Browser History Injection               \\ \hline
5                                                           & Fake Flash Player Download                                                                 & JSCoinminer Download         & Mass Injection Website                       \\ \hline
\end{tabular}
}
\caption{Top 5 events with most usage changes in selective dates. }
\label{tab:top_changes}
\end{table*}

\noindent \textbf{Global change detection.} Recall that temporal event embeddings are the solution of a joint optimization problem across all time slices (see Section~\ref{sec:methodology}). 
Therefore, such embeddings not only encode their respective usage and context in a given time slice, but also its history across all time slices. In this section, we show how to leverage the temporal embeddings generated by \approach to find times where we observe an anomalous high number of changes in the use of multiple security events.

\approach computes a list of changes for each security event $e_i$ in all time slices using the weighted drift metric (see Eq~\ref{eq:changepoint}). 
We use this to identify the observation periods in which many security events exhibit most usage changes. 
These points are interesting candidates for scrutiny for security analysts, since multiple changes in word embeddings might indicate the emergence of new pervasive attacks.
Note that we remove all the security events with less than 100 observations in 2 years time (see Section~\ref{sec:discussion} for a rationale for this). Figure~\ref{fig:changepoints} shows a histogram of the time slices in which security event usage changes most. As we can see,  we observe 34 security events with most usage changes between October 11 and October 18 2018. We selectively list the top 5 events with most usage changes in different dates (see Table~\ref{tab:top_changes}). These changes demonstrate how security event evolve over the time. For example, at the beginning of 2017, we can observe many changes relating to exploits affecting routers.
This can be an indicator of a large attack campaign targeting such devices unfolding.
Across time, more attacks change over fake tech support websites, coinminer and content management systems. 
This context information is usually for analysts to improve situational awareness and be promptly warned about emerging attacks.

\subsection{Trend Identification}
\label{sec:robustness}

\begin{figure}[t]
    \centering
    \begin{subfigure}{0.95\linewidth}
        \centering
        \includegraphics[width=\linewidth]{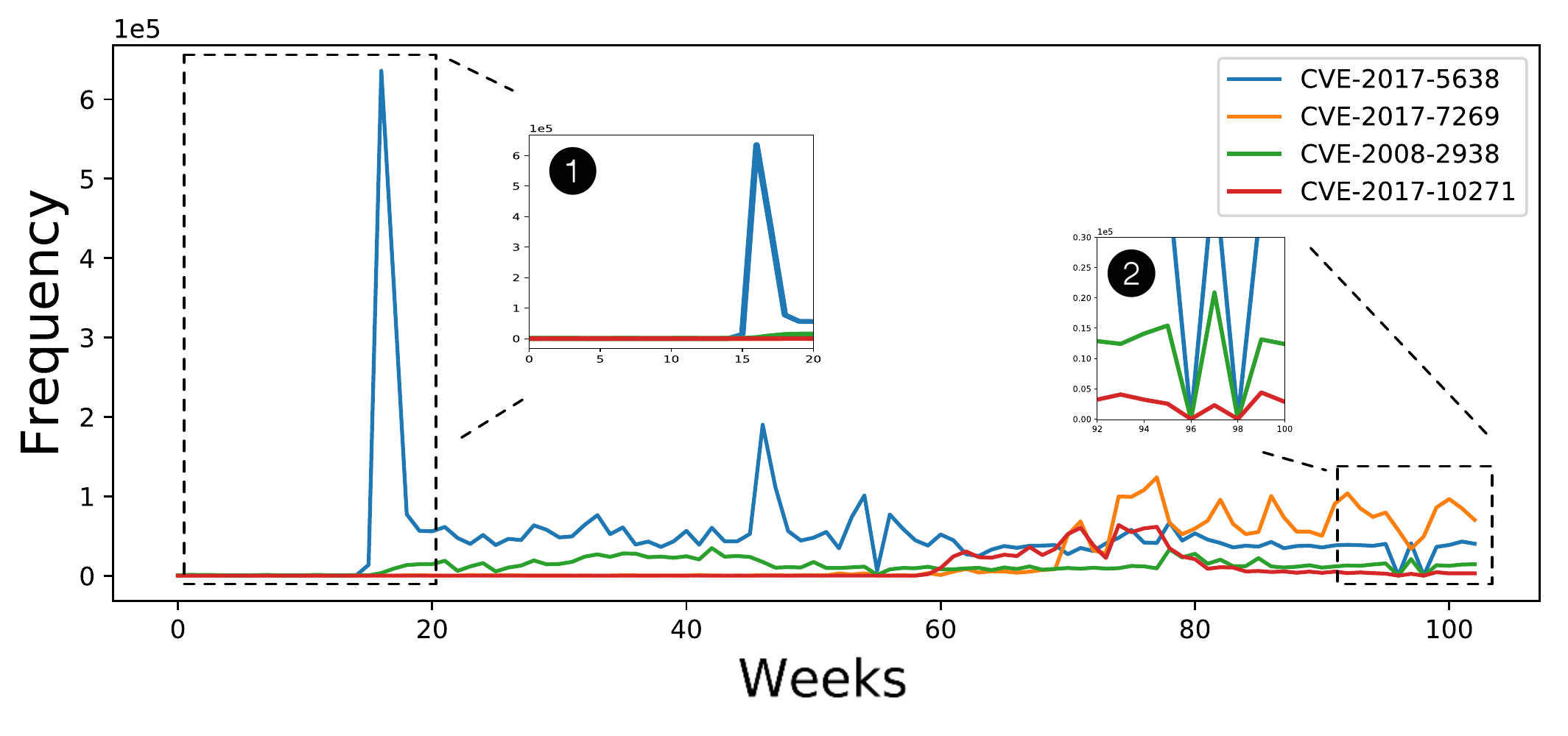}
        \caption{}
        \label{fig:populairty_f}
    \end{subfigure}
    \hfill
    \begin{subfigure}{0.95\linewidth}
        \centering
        \includegraphics[width=\linewidth]{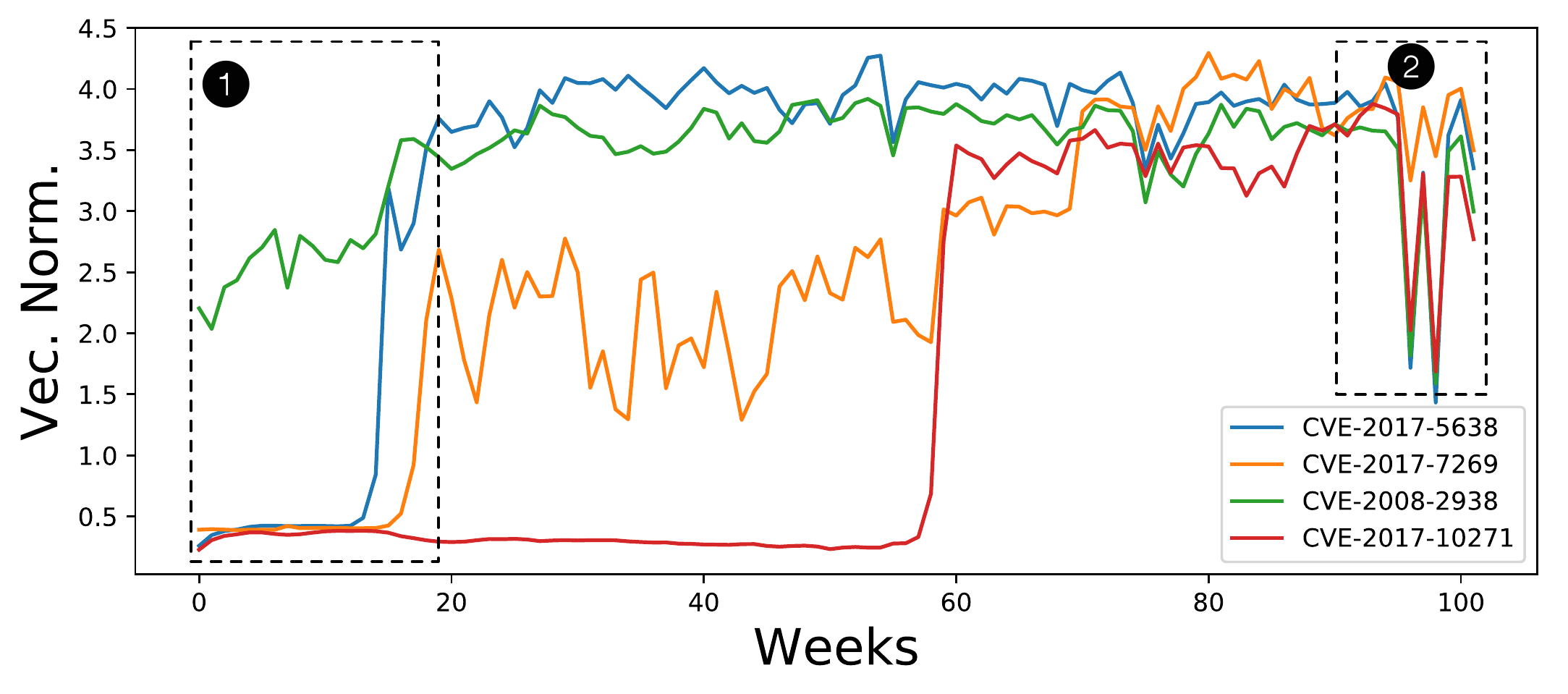}
        \caption{}
        \label{fig:populairty_e}
    \end{subfigure}
	\caption{Event frequencies (\ref{fig:populairty_f}) and event embedding vector norms (\ref{fig:populairty_e}). We select four CVEs relating to different Web services to demonstrate the robustness of temporal event embedding in trend changes (CVE-2008-2938 in green line, CVE-2017-5638 in blue line, CVE-2017-10271 in red line, and CVE-2017-7269 in orange line). The dashed box highlights the observation period discussed in Section~\ref{sec:robustness}.} 
	\label{fig:robustness} 
\end{figure}

{One \emph{de facto} method used in empirical studies~\cite{bilge2012before, nappa2015attack, vervier2015mind, li2017large} to analyze temporal usage changes is leveraging frequencies to reveal patterns.
That is, previous approaches often start by determining the occurrence frequency of events across the data, and using the event frequency time series as a criterion to reveal the significance of these events. Despite its straightforwardness in analyzing and visualizing temporal data, one drawback of this approach is that frequencies are prone to noise because they are only counted at a specific timestamp.
Another drawback of using temporal frequency, especially in practical analysis, is that it is more difficult to compare two time series when they are both trending but at a different magnitude, and a sudden spike in the occurrence of one security event might make it difficult to identify other important events that are not happening as frequently. 
To demonstrate this problem, we use four popular remote Web server attack vectors (see Figure~\ref{fig:populairty_f}). We can observe in Figure~\ref{fig:populairty_f} (inset \mbox{\ding{182}}) , that Apache Struts CVE-2017-5638 (blue line) was overwhelmingly exploited by attackers after it was disclosed. Its preponderant usage overshadows (see the zoom region inset \mbox{\ding{182}} in Figure~\ref{fig:populairty_f}) the other popular remote Web server attack vectors (e.g.,  HTTP Apache Tomcat UTF-8 Dir Traversal CVE-2008-2938) when using event frequency time series to comparatively study attack vector popularity.

\begin{figure}[t]
    \centering
    \begin{subfigure}{0.95\linewidth}
        \centering
        \includegraphics[width=\linewidth]{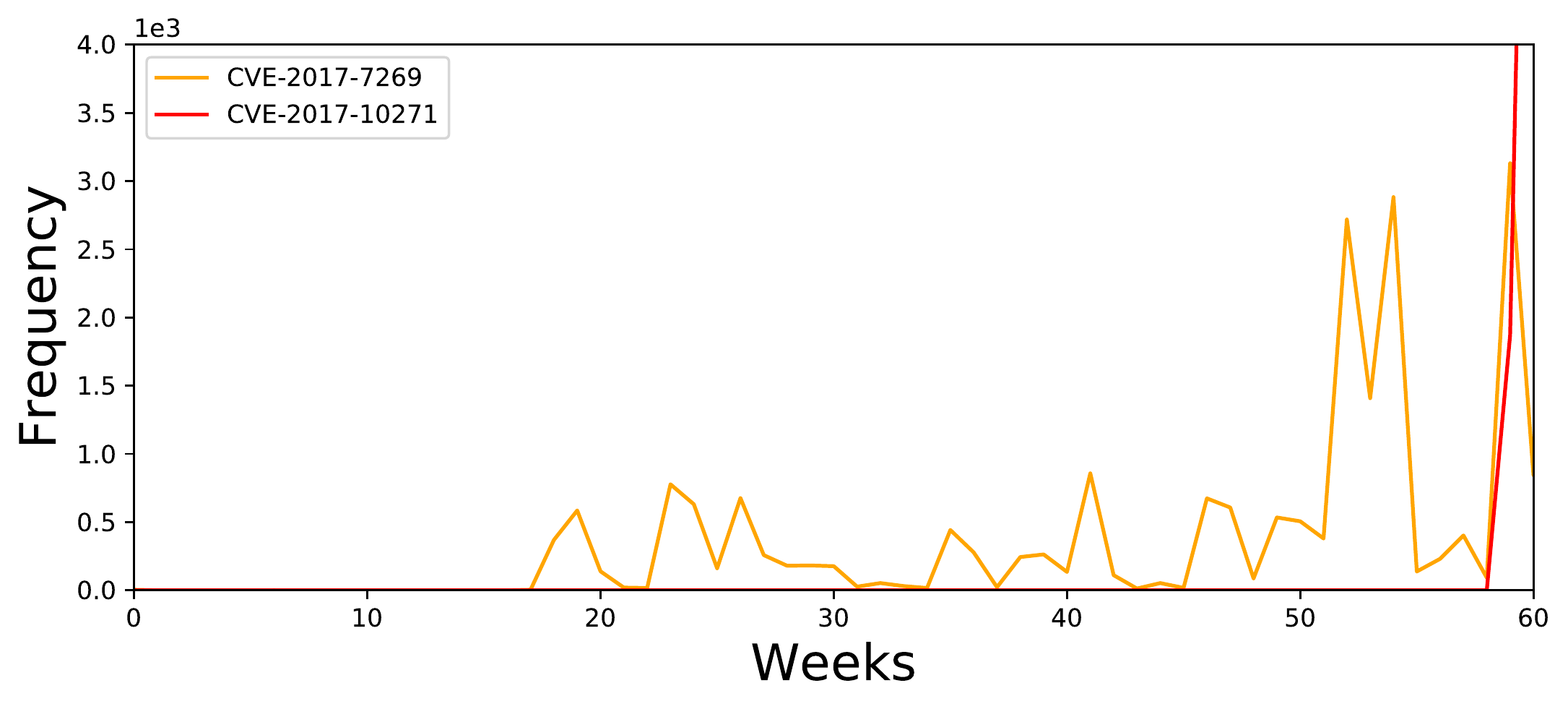}
        \caption{}
        \label{fig:populairty_f_special}
    \end{subfigure}
    \hfill
    \begin{subfigure}{0.95\linewidth}
        \centering
        \includegraphics[width=\linewidth]{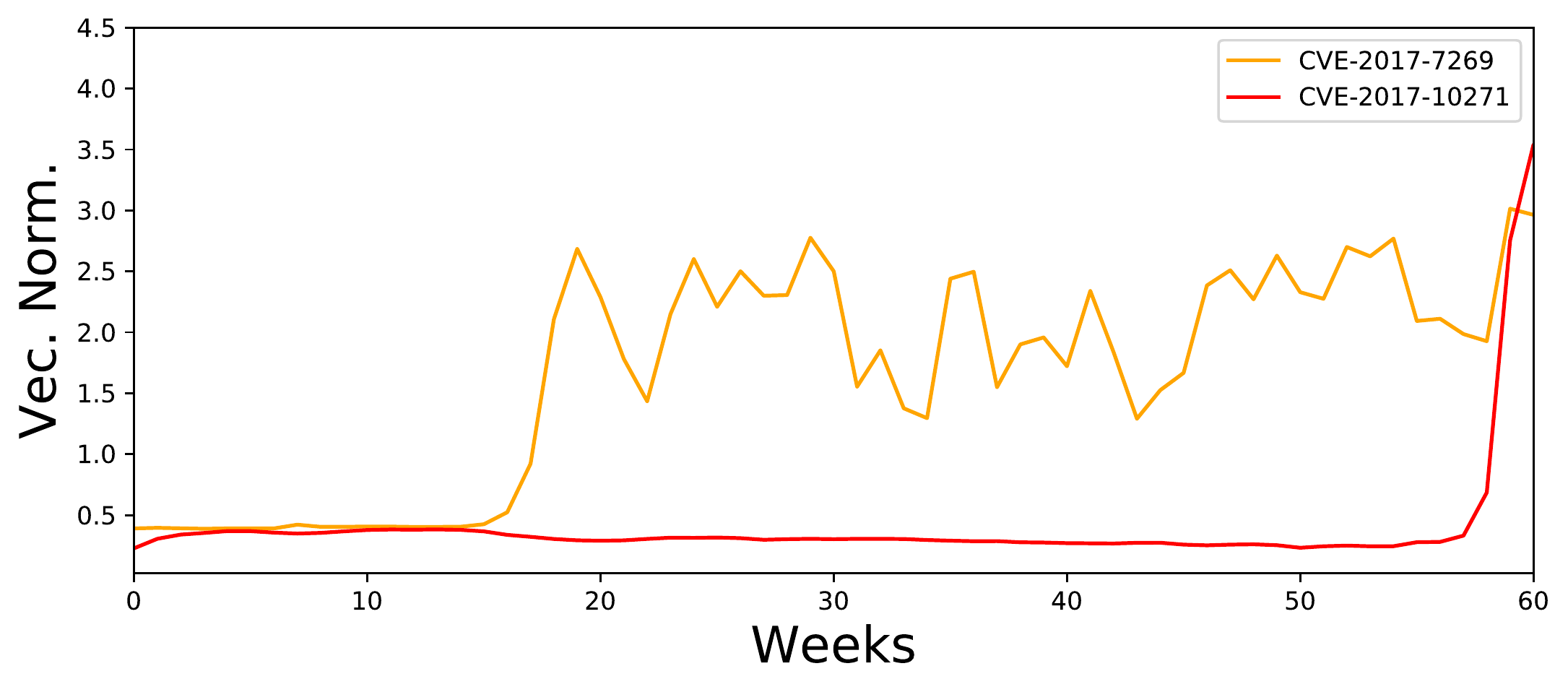}
        \caption{}
        \label{fig:populairty_e_special}
    \end{subfigure}
	\caption{Comparison study of CVE-2017-7269 (orange line) and CVE-2017-10271 (red line). \mbox{\approach} can trace the emerging popularity of CVE-2017-7269 even though it is barely observable in Figure~\ref{fig:robustness}. } 
	\label{fig:robustness_special} 
\end{figure}

In the rest of this section, we demonstrate that the word embeddings calculated by \approach are more robust than temporal frequencies to reveal trend changes. Recall the definition of $PMI_t(c, \mathcal{S})$ (defined in Section~\ref{sec:methodology}). It contains information from $W(e_i, e_j)$, the number of times $e_i$ and $e_j$ co-occurring in a given contextual window, and $W(e_i)$ and $W(e_j)$ respectively count the occurrences of security events $e_i$ and $e_j$. The norms of word embeddings $\eta_{e_i}$, as a consequence of matrix factorization of $PMI_t(c, \mathcal{S})$, grow with word frequency and are averaged by their contexts. Therefore, we can leverage these norms to identify event usage trends. 
We compare temporal frequencies (Figure~\ref{fig:populairty_f}) to the embedding vector norm (Figure~\ref{fig:populairty_e}) in the context of trend identification. It is straightforward to see that the security event ``Apache Struts CVE-2017-5638'' was predominantly leveraged by the attackers for a short period of time. As we said, such sudden spike makes other trendy CVEs used at that same period of time less detectable (Figure ~\ref{fig:populairty_f}). However, we can see that the temporal event embeddings (Figure ~\ref{fig:populairty_e}) reveal the real patterns behind the frequencies, and are able to capture trends more reliably. 
For example, we can clearly see in Figure~\ref{fig:populairty_e} that  CVE-2008-2938 (``HTTP Apache Tomcat UTF-8 Dir Traversal,'' green line in Figure ~\ref{fig:populairty_f}) 
is a persistent  vector used by the attackers. 
This was not observable in Figure~\ref{fig:populairty_f} because CVE-2017-5638 dominates during that period of time. Nevertheless, it is worth noting that we study the independent popularity of these attack vectors and reveal their underlying usage trend beneath the event frequency time series. Hence in this paper we do not intend to study how such attack vectors influence each other. It is coincidental that the red (CVE-2017-10271) and green (CVE-2008-2938) lines show a similar vector norm shift. In fact, these vector norm shifts were incurred by their similar usage changes (\eg dropping close to zero) during the same period (see inset \mbox{\ding{183}}, Figure ~\ref{fig:populairty_f}). \mbox{\approach} still preserves the trends of red (CVE-2017-10271) and green (CVE-2008-2938) lines (\ie larger than zero) in this extreme case and their respective vector norms become stable immediately after.

Our temporal embedding can also capture how two CVEs, CVE-2017-10271 (``Oracle WebLogic RCE,'' red line in Figure~\ref{fig:populairty_f})
and CVE-2017-7269 (``Buffer overflow in the ScStoragePathFromUrl function in the WebDAV service in IIS 6.0,'' orange line in Figure ~\ref{fig:populairty_f}) 
gradually emerge as a trendy attack vector over time. That is, both CVE-2017-10271 (red line) and CVE-2017-7269 (orange line), before their disclosure dates (May 27, 2017 and Oct 17, 2017), are flat. This proves that \mbox{\approach} faithfully captures their non-existent trend. After their disclosure, \mbox{\approach} is able to correctly keep track of their trends. For example, in Figure~\ref{fig:robustness_special} we show that our approach  can reliably trace the trends of CVE-2017-7269 (orange line) from approximately week 19 in the figure.  Although its usage (e.g., frequency) is less observable than CVE-2008-2938 and CVE-2017-5638 during the same period of time (see  Figure~\ref{fig:robustness}), its emerging trend is still recognized by the increase in vector norm as we can see in Figure~\ref{fig:populairty_e_special}. Additionally, despite the five times of frequency difference between it initial disclosure (week 19 in Figure~\ref{fig:populairty_f_special}) and later weeks (after week 51 in Figure~\ref{fig:populairty_f_special}), \mbox{\approach} captures the trend changes with a small fluctuation of less than 0.6. \mbox{\approach} also reliably captures CVE-2017-10271 (red line) non-existent trend before its public disclosure.

\subsection{Event Evolution}
\label{sec:case1}

\begin{table}[t]
\centering
\resizebox{\linewidth}{!} {
\begin{tabular}{|l|l|}
\hline
\multicolumn{2}{|c|}{\textbf{Drupal core RCE (CVE-2018-7602)}}                                                                                                                           \\ \hline
\multicolumn{1}{|c|}{May 15 2018}                                                             & \multicolumn{1}{c|}{Nov. 08 2018}                                               \\ \hline
Joomla JCE Vulnerability                                                                      & phpMyAdmin RFI (CVE-2018-12613)                                                 \\ \hline
\begin{tabular}[c]{@{}l@{}}Wordpress RevSlider/ShowBiz Bypass \\ (CVE-2014-9735)\end{tabular} & \begin{tabular}[c]{@{}l@{}}Drupal SQL Injection \\ (CVE-2014-3704)\end{tabular} \\ \hline
WordPress Symposium Plugin Shell Upload                                                       & Adobe Flex BlazeDS RCE (CVE-2017-3066)                                          \\ \hline
\end{tabular}
}
\caption{Top 3 security events associated with CVE-2018-7602 at the beginning (May 15 2018) and the end (November 18 2018) of our time span.}
\label{tab:30762_casestudy}
\end{table}

\begin{figure*}
    \centering
    \begin{subfigure}{0.4\linewidth}
        \centering
        \includegraphics[width=\linewidth]{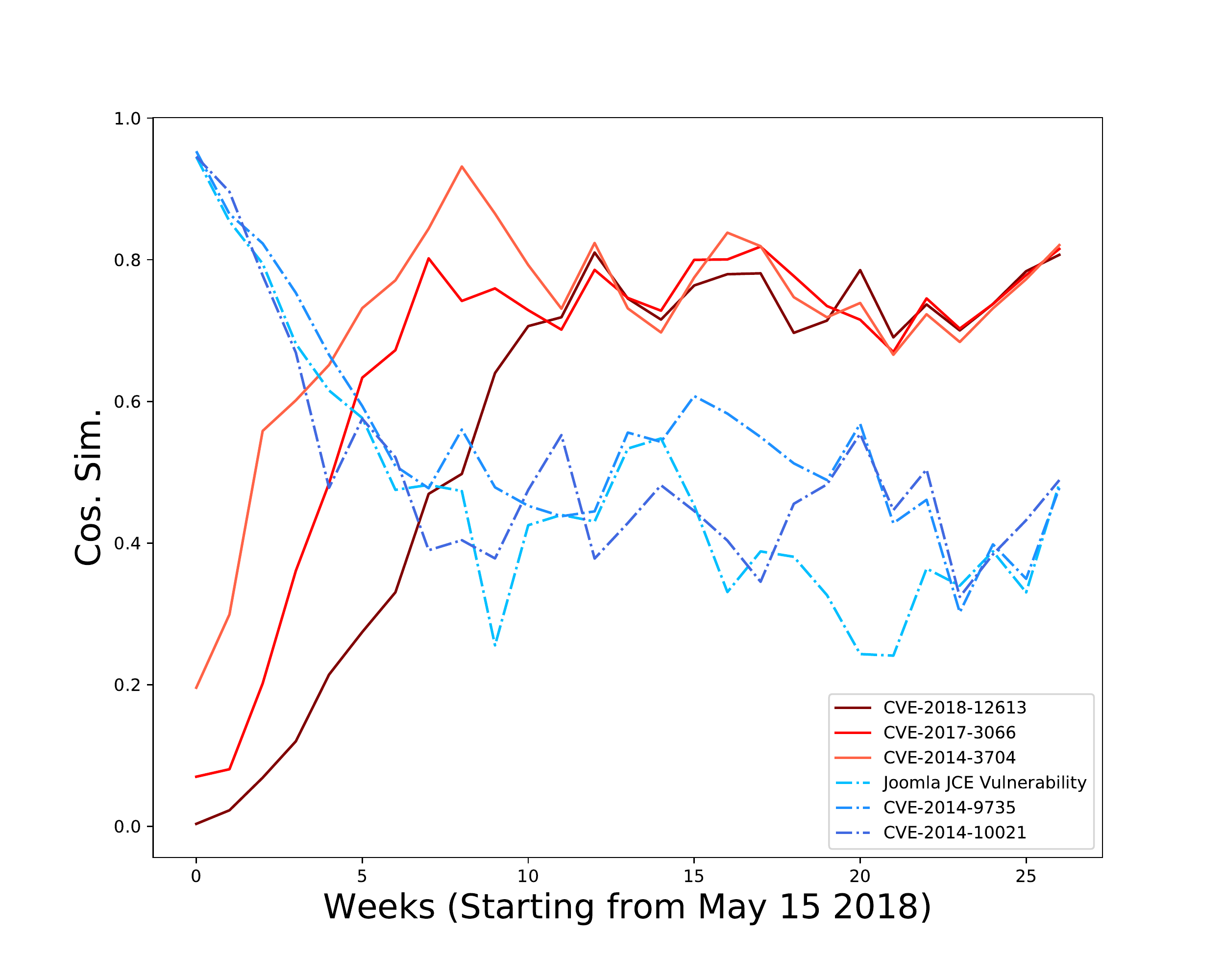}
        \caption{Joomla JCE Vulnerability, CVE-2014-9735, CVE-2014-10021 (blue dashed lines) and CVE-2018-12613, CVE-2017-3066, CVE-2014-3704 (red solid lines) respectively are the top three closest security events associated with Drupal core RCE (CVE-2018-7602) at the beginning (end) of the observation span (starting from May 15 2018).}
        \label{fig:30762_up_down}
    \end{subfigure}
    \quad
    \begin{subfigure}{0.4\linewidth}
        \centering
        \includegraphics[width=\linewidth]{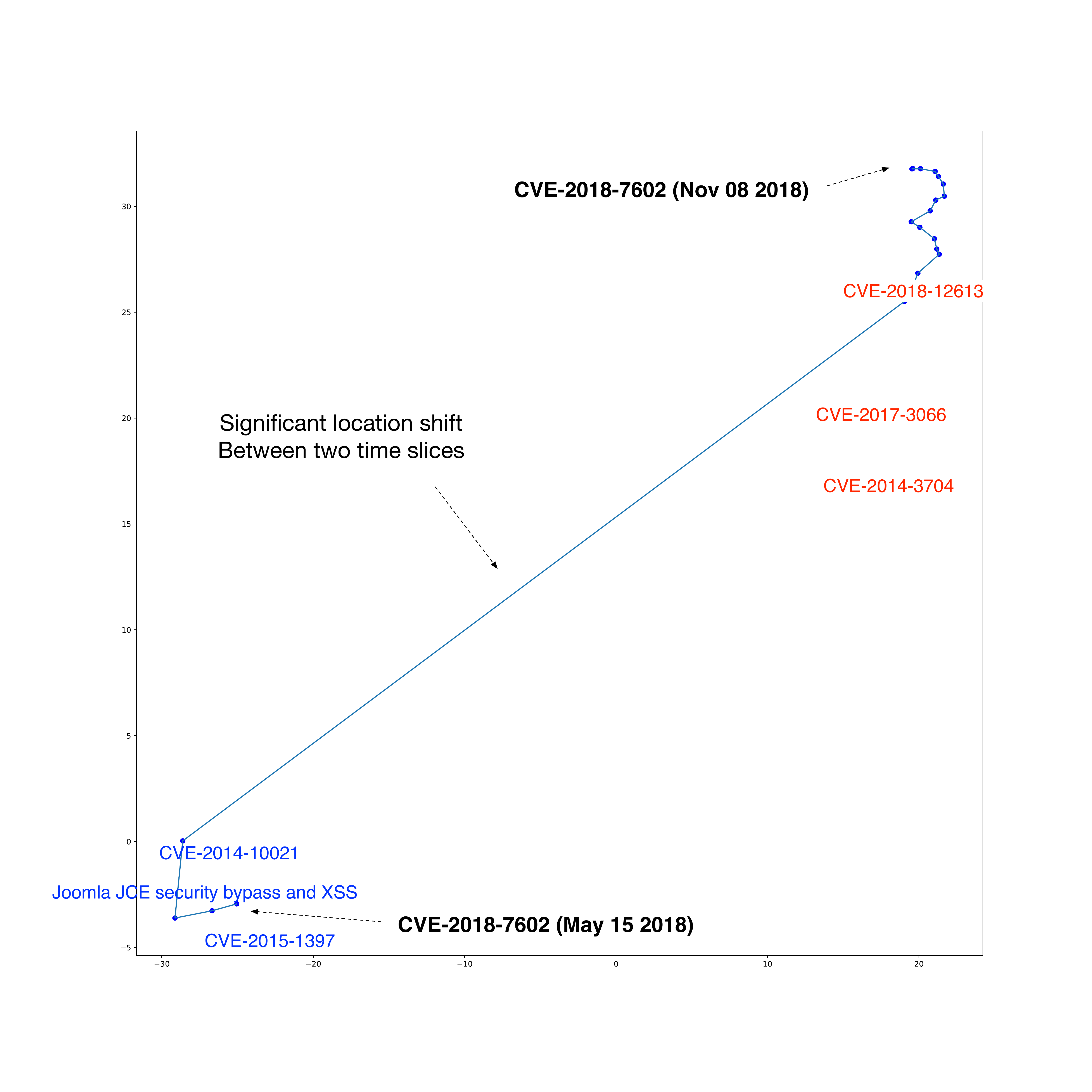}
	\caption{Temporal t-SNE trajectory of Drupal core RCE (CVE-2018-7602). The significant shift in this figure can be indicative of the old reconnaissance campaign fading and of the new one more targeted towards Drupal emerging, or of an attacker changing their behavior.}
        \label{fig:30762_tsne}
    \end{subfigure}
    
	\caption{Drupal core RCE (CVE-2018-7602) Evolution between May 15 2018 and November 8 2018.}
	\label{fig:30762_casestudy} 
\end{figure*}

Another useful functionality for which \approach can be used is understanding how attacks evolve in the wild, and in particular monitoring which attack steps are often performed together by attackers.
In a nutshell, security events that are often used together will have similar contexts.
Identifying events with such similar contexts could help detecting emerging threats such as new botnets scanning for specific vulnerabilities (\eg Mirai or WannaCry) or new exploit kits that are probing for specific weaknesses in victim systems~\cite{nayak2014some, antonakakis2017understanding}.

To evaluate \approach's capability of tracking the evolution of security event contexts over time in relation to each other we use the same CVE from Section~\ref{sec:motivation}, ``Drupal core RCE (CVE-2018-7602).'' 
This CVE is for a highly critical remote code execution (RCE) vulnerability that exists within multiple subsystems of Drupal 7.x and 8.x, enabling attackers to compromise a machine running a Drupal website. This vulnerability was first disclosed on April 23 2018, and we observe its activities in the our data starting from May 15 2018. Due to its high severity, it is interesting to see how such a critical vulnerability was exploited in different contexts across time. 

We use Eq~\ref{equ:neighbor} to identify the top 3 security events associated with CVE-2018-7602 at the beginning and the end of our time span.
These events are the ones that have the closest context to CVE-2018-7602, and are therefore used in association with it. 
Table~\ref{tab:30762_casestudy} shows a detailed description of these events. 
It is interesting to note that the top 3 attack vectors used together with CVE-2018-7602 at the beginning resemble a reconnaissance attack aiming at all three major content management systems - Joomla, Wordpress, and Drupal. 
Toward the end of the time span, however, we notice that CVE-2018-7602 migrates to be part of a more specific multi-step attack, aiming at the ecosystem surrounding the Drupal CMS - php (phpMyAdmin), SQL (Drupal SQL Injection), and Flex (BlazeDS RCE). 

Leveraging Eq~\ref{equ:cosine}, we show in Figure~\ref{fig:30762_up_down} that the temporal security event embedding computed by \approach can meaningfully capture the aforementioned usage changes across time. ``Joomla JCE Vulnerability,'' CVE-2014-9735, CVE-2014-10021 (blue dashed lines), and CVE-2018-12613, CVE-2017-3066, CVE-2014-3704 (red solid lines) respectively are the top three closest security events associated with ``Drupal core RCE (CVE-2018-7602)'' at the beginning (end) of the observation span (starting from May 15 2018). We can clearly see that the red lines are rising (\ie these security events are used more closely with CVE-2018-7602), and the blue lines are moving away from CVE-2018-7602. In general, we can see that the attackers change their modus operandi, using CVE-2018-7602 as part of a more targeted attack on Drupal, 8 weeks after its initial observation in the telemetry data. 
This can be indicative of the old reconnaissance campaign fading and of the new one more targeted towards Drupal emerging, or of an attacker changing their behavior.
Note that such usage changes can be automatically detected using various change point detection algorithms~\cite{aminikhanghahi2017survey}. 

\noindent \textbf{Trajectory visualization}. The trajectory of a security event in the embedded latent space can assist security analysts to understand its context changes over time. To show this, we collect the top $k$ security events associated with CVE-2018-7602 using Eq~\ref{equ:neighbor} in each time slice to form our trajectory data $D_{e_i} = \{\mathcal{N}_k(e_i^{(t)})\}$,\noindent where $t$ starts from May 15, 2018. We accordingly plot the 2-D t-SNE projection of the temporal embeddings of CVE-2018-7602 (and the aforementioned security events) in Figure~\ref{fig:30762_tsne} to visualize its context change over time. In Figure~\ref{fig:30762_tsne}, each blue dot represents the 2-D location in the latent space at a given timestamp. We can observe a considerable drift in Figure~\ref{fig:30762_tsne} between the fourth and the fifth blue dot. This is correlated to the trend we observed in Figure~\ref{fig:30762_up_down}. The top 3 security events associated with CVE-2018-7602 at the beginning and the end of our time span are closer to the respectively locations in Figure~\ref{fig:30762_tsne}. Note that such changes in Figure~\ref{fig:30762_tsne} can be quantitatively detected as these locations are bounded in a Euclidean space.

\subsection{System Performance}
\label{sec:performance}
\mbox{\approach} is implemented in Python 3.7.3 and tested on a server with dual Xeon E5-2630 CPUs and 256GB memory running Ubuntu Linux 14.04. In this setup, \mbox{\approach} takes 859.86 seconds to construct the PPMI matrices for all 102 snapshots. Once the PPMI matrices are constructed, \mbox{\approach} takes 3014.18 seconds per epoch to optimize the temporal embeddings (see Section~\ref{sec:methodology}). We empirically run 5 epochs for \mbox{\approach} to reach the optimum embedding results. This leads to approximately 4.18 hours for \mbox{\approach} to generate final temporal embeddings for all 8,087 security events across 102 snapshots. This enables us to deploy \mbox{\approach} to understand the long term evolution of different security events at scale by security analysts.

\subsection{End-to-end Evaluation of \approach}
\label{sec:evolution}

\label{sec:case_study}
\begin{table}[t]
\resizebox{\linewidth}{!} {
\begin{tabular}{|l|l|}
\hline
\multicolumn{2}{|c|}{\textbf{Apache Struts Jakarta Multipart parser RCE (CVE-2017-5638)}} \\ \hline
\multicolumn{1}{|c|}{Mar. 23 2017}          & \multicolumn{1}{c|}{Nov. 08 2018}           \\ \hline
WifiCam Authentication Bypass               & Malicious OGNL Expression Upload            \\ \hline
CCTV-DVR Remote Code Execution              & Apache Struts CVE-2017-12611                \\ \hline
ZyNOS Information Disclosure                & Malicious Serialized Object Upload          \\ \hline
\end{tabular}
}
\caption{Top 3 security events associated with CVE-2017-5638 at the beginning (March 9 2017) and the end (November 8 2018) of our time span.}
\label{tab:29972_casestudy}
\vspace{-0.2cm}
\end{table}

\begin{figure}
    \centering
    \begin{subfigure}{0.92\linewidth}
        \centering
        \includegraphics[width=\linewidth]{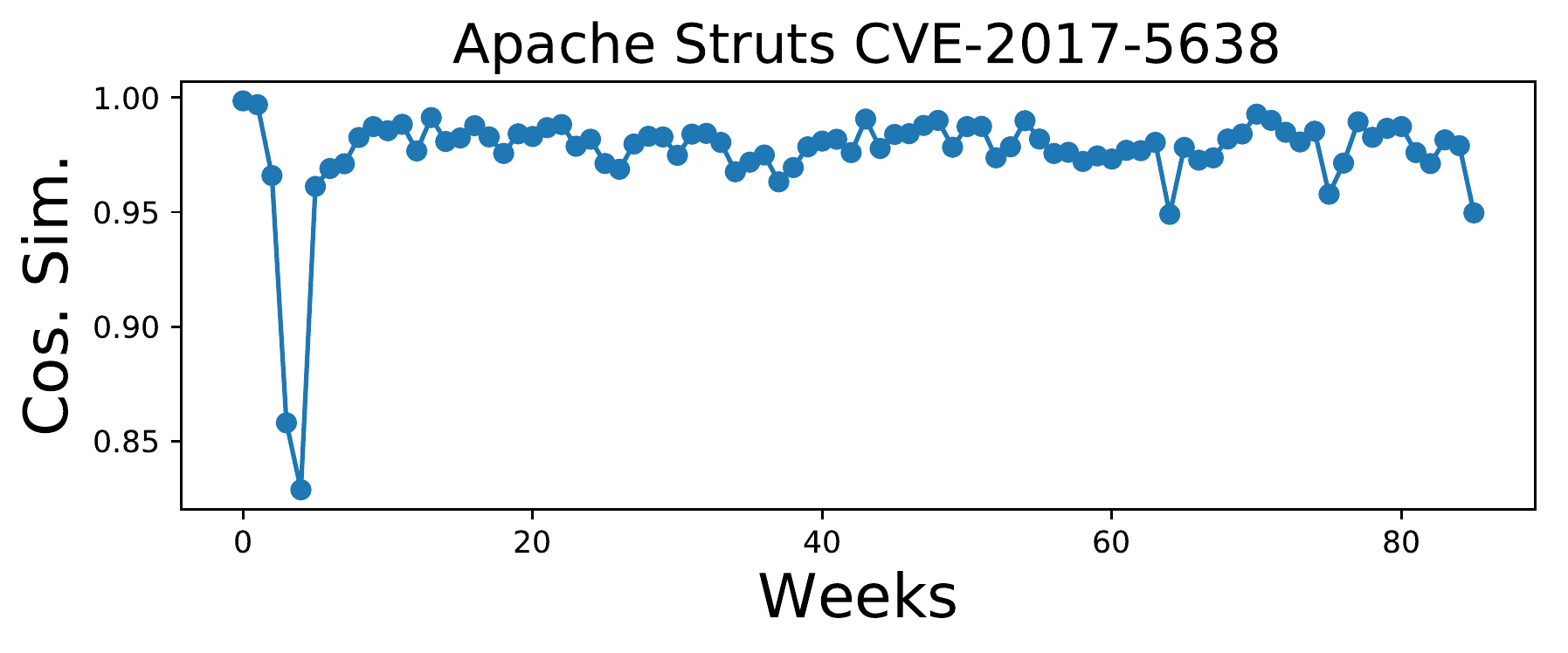}
	\caption{Temporal embedding result of Apache Struts Jakarta Multipart parser RCE  (CVE-2017-5638). Cosine similarity values for each plot is calculated as $similarity(\eta_{e_i}^{(t-1)}, \eta_{e_i}^{(t)})$, where $t$ starts from March 9 2017.}
        \label{fig:29972_dynamic}
    \end{subfigure}
	\hfill
    \begin{subfigure}{0.8\linewidth}
        \centering
        \includegraphics[width=\linewidth]{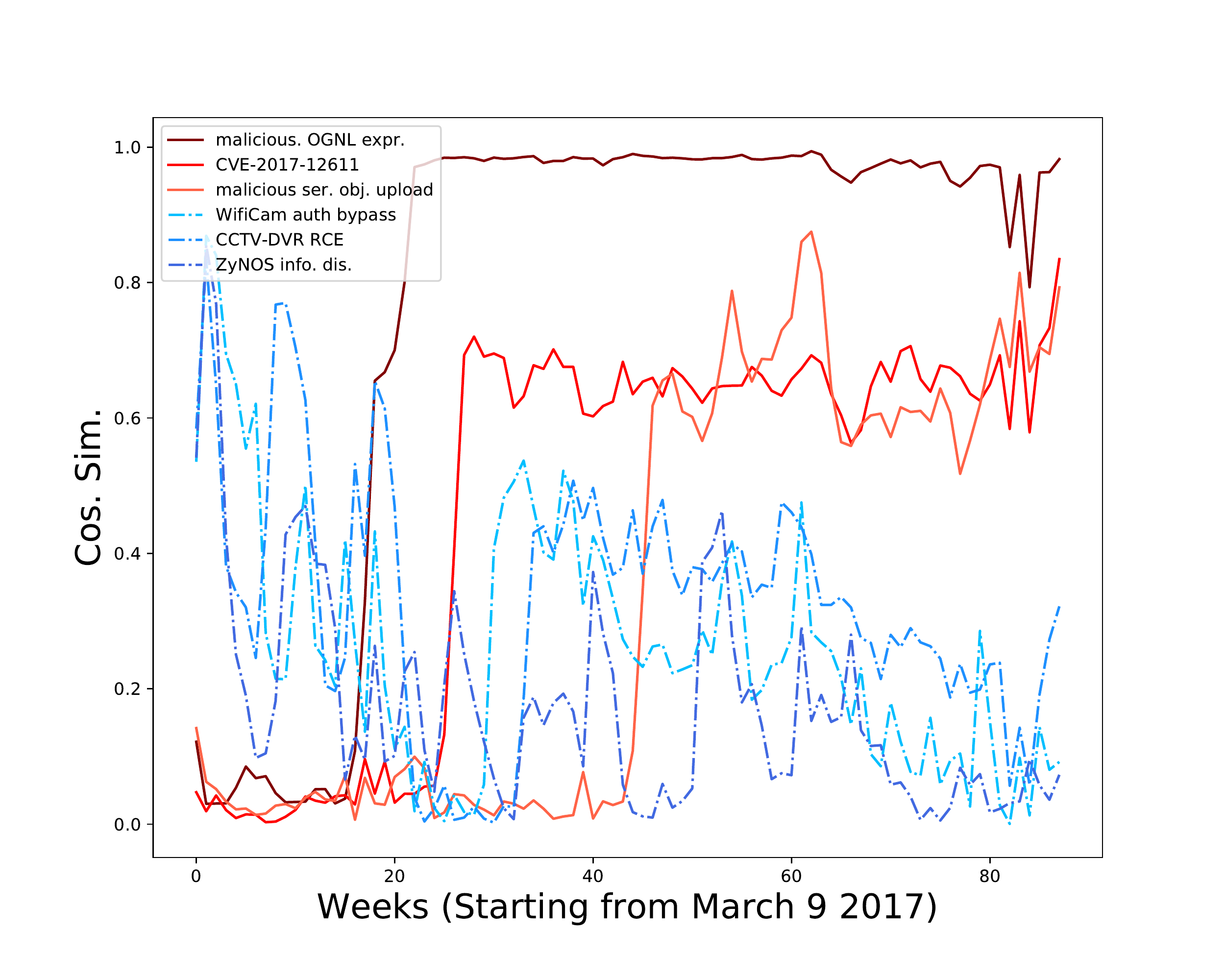}
        \caption{WifiCam Authentication Bypass, CCTV-DVR Remote Code Execution, ZyNOS Information Disclosure (blue dashed lines) and  Malicious OGNL expression upload, CVE-2017-12611 and Malicious Serialized Object Upload (red solid lines) are the top three closest security events associated with CVE-2017-5638 at the beginning (end) of the observation span (starting from March 9 2017), respectively.}
        \label{fig:29972_up_down}
    \end{subfigure}
    \hfill
    \begin{subfigure}{0.8\linewidth}
        \centering
        \includegraphics[width=\linewidth]{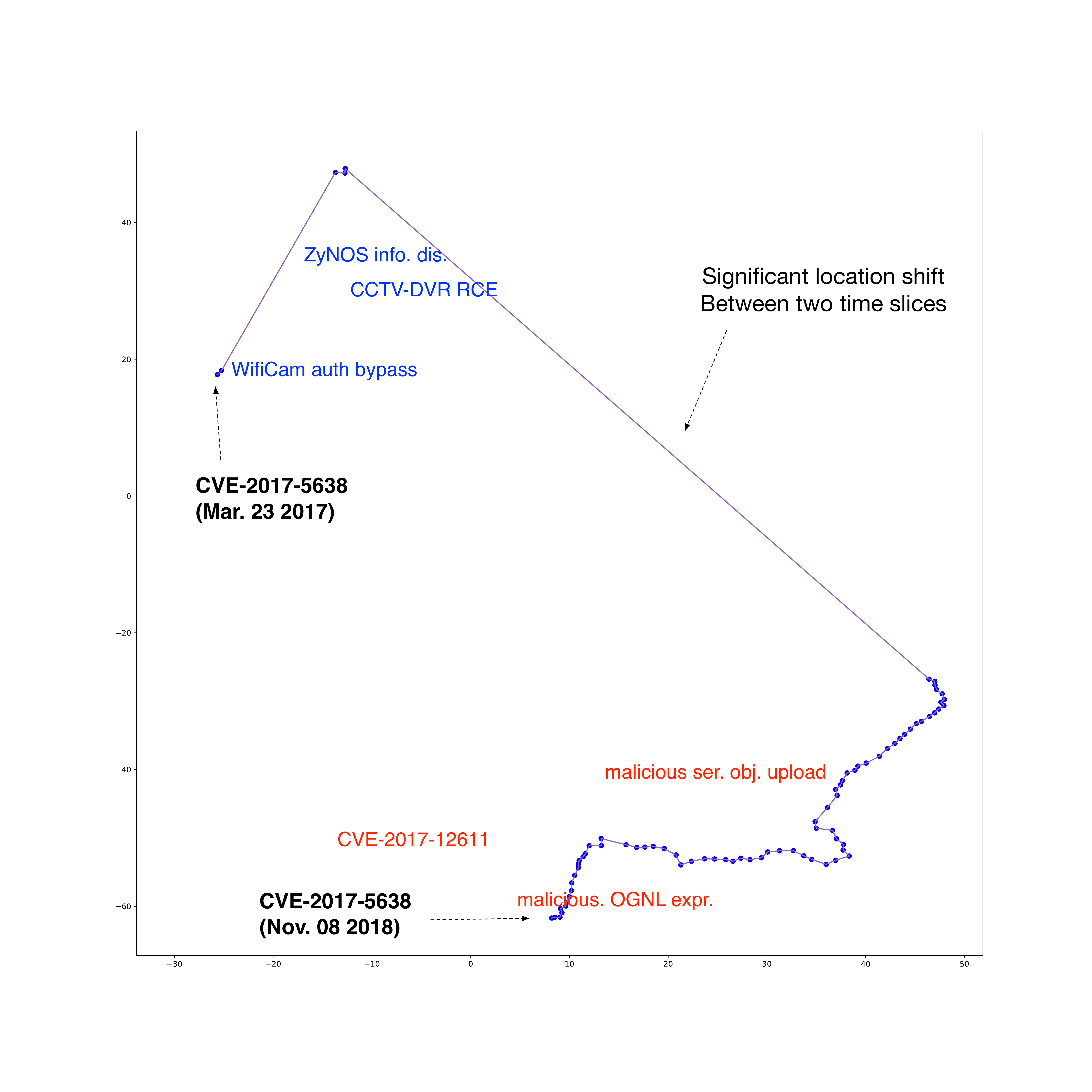}
        \caption{Temporal t-SNE trajectory of Apache Struts Jakarta Multipart parser RCE  (CVE-2017-5638).}
        \label{fig:29972_tsne}
    \end{subfigure}
	\caption{Apache Struts Jakarta Multi- part parser RCE (CVE-2017-5638) evolution between March 9 2017 and November 8 2018.}
	\label{fig:29972_casestudy} 
\end{figure}

In the previous sections we evaluated the ability of \approach to study various aspects of how a security event is exploited in the wild.
We envision that \mbox{\approach} could be used by security analysts to understand the evolution of the use of a security event (e.g., a vulnerability) over time.
In this section we provide an end-to-end example to show how an analyst could be using our tool to better understanding the context surrounding a security vulnerability and its evolution.
To this end, we study the evolution of the security event ``Apache Struts Jakarta Content-Type RCE (CVE-2017-5638),'' a remote code execution vulnerability targeting Apache Struts.
This vulnerability is classified by NVD as a critical bug with CVSS score 10.0, and was the culprit of the Equifax data breach~\cite{apache2017media}.

Figure~\ref{fig:29972_dynamic} shows that CVE-2017-5638 experienced a change in the way it is being exploited between April 5, 2017 and April 13, 2017.
\mbox{\approach} is able to identify this change in the way the vulnerability is exploited, and an analyst could easily identify this.
We then want to understand what these two contexts looked like, and evaluate whether this information can help us understand the types of attacks that CVE-2017-5638 was used in.
To this end we perform the analysis described in Section~\ref{sec:case1}, whose results are shown in Figure~\ref{fig:29972_up_down}.
In particular, we trace the top three security events with the closest context to CVE-2017-5638 at the beginning and at the end of our analysis period (see Table~\ref{tab:29972_casestudy} for a detailed description of these events).
The top 3 security events tightly associated with CVE-2017-5638 at the beginning of our observation span are IoT specific attack vectors.
Figure~\ref{fig:29972_up_down} shows that these three IoT attack vectors maintain similar contexts to CVE-2017-5638 for approximately 10 weeks, indicating that the four vulnerabilities were frequently exploited together in the wild as part of a multi-step attack.
Later in the analysis period (starting from May 8, 2017) we see that this attack is substituted by another attack that is targeted at the Apache Struts ecosystem, consisting of Malicious OGNL expression upload, CVE-2017-12611 and Malicious Serialized Object Upload.
Figure~\ref{fig:29972_tsne} shows a similar pattern, with CVE-2017-5638 migrating from being close to the IoT security events to the Apache Struts related ones.
This information could inform an analyst about the change in which CVE-2017-5638 was exploited in the wild, switching from an attack step as part of an IoT-centered attack to part of an attack centered around Apache Struts.

To further evaluate the accuracy of the embeddings calculated by \mbox{\approach}, and the meaningfulness of our analysis, we further investigate the contexts calculated at the beginning of our analysis period, when
CVE-2017-5638 was exploited in conjunction with IoT-related vulnerabilities.
To this end, we retrieve the remote IP addresses from which the security events originated on selected dates. 
We find that in many cases these four (including CVE-2017-5638) security events were generated from connections originating from the same IP addresses (for example, 701 unique IP addresses on March 23 2018), indicating that the relation depicted by the context is not an artifact of \mbox{\approach}, but it is indeed a large scale attack performed by the same malicious actors.
These IP addresses were often located in residential ISPs, indicating a potential botnet infection.

We later found confirmation that a variant of the Mirai IoT malware was active during that period and was explicitly exploiting CVE-2017-5638, ``WifiCam auth bypass,'' ``CCTV-DVR RCE,'' and ``ZyNOS information disclosure,'' the same vulnerabilities picked up by \approach as being related (see Table~\ref{tab:29972_casestudy})~\cite{unit422018mirai}.
This shows that \approach can help identifying complex attacks in an effective way.
An additional advantage of our approach is that our system was able to flag this attack as emerging in the wild 72 weeks before it was actually discussed by security researchers, highlighting the potential of the use of temporal embeddings for early warning and situational awareness.

\section{Limitations and Discussion}
\label{sec:discussion}
\begin{figure*}[t]
    \centering
    \begin{subfigure}{0.3\linewidth}
        \centering
        \includegraphics[width=\linewidth]{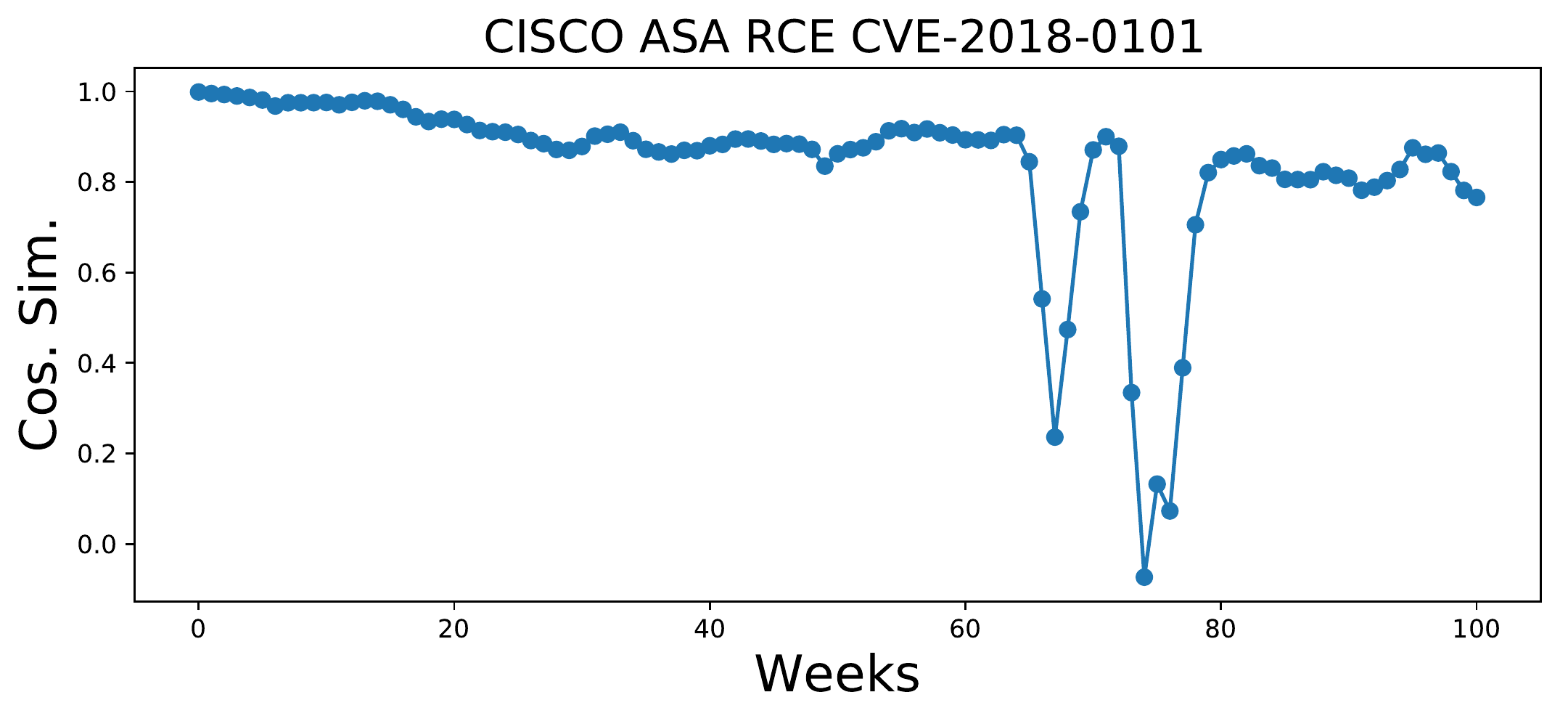}
        \caption{}
        \label{fig:30627_stable}
    \end{subfigure}
    \quad
    \begin{subfigure}{0.3\linewidth}
        \centering
        \includegraphics[width=\linewidth]{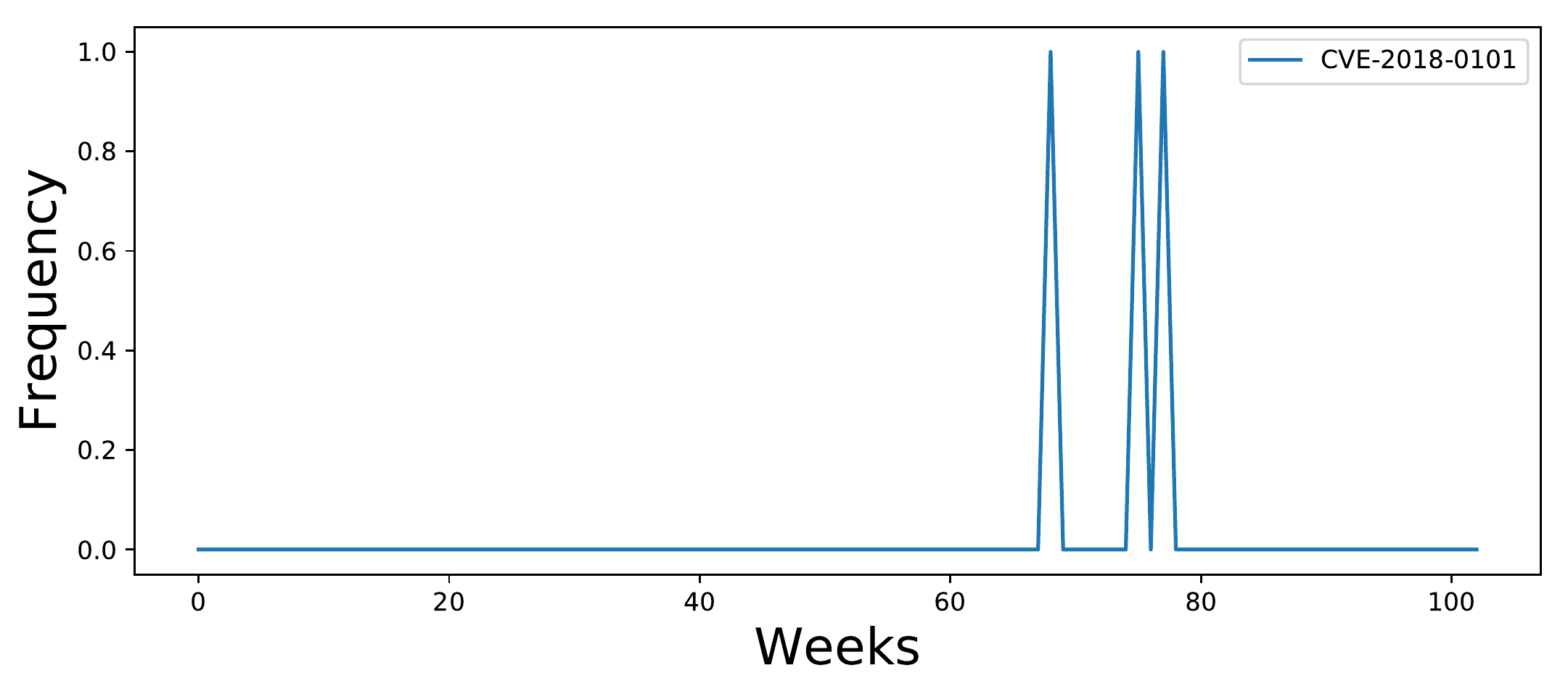}
        \caption{}
        \label{fig:fig:popularity_f_30627}
    \end{subfigure}
	\quad
    \begin{subfigure}{0.3\linewidth}
        \centering
        \includegraphics[width=\linewidth]{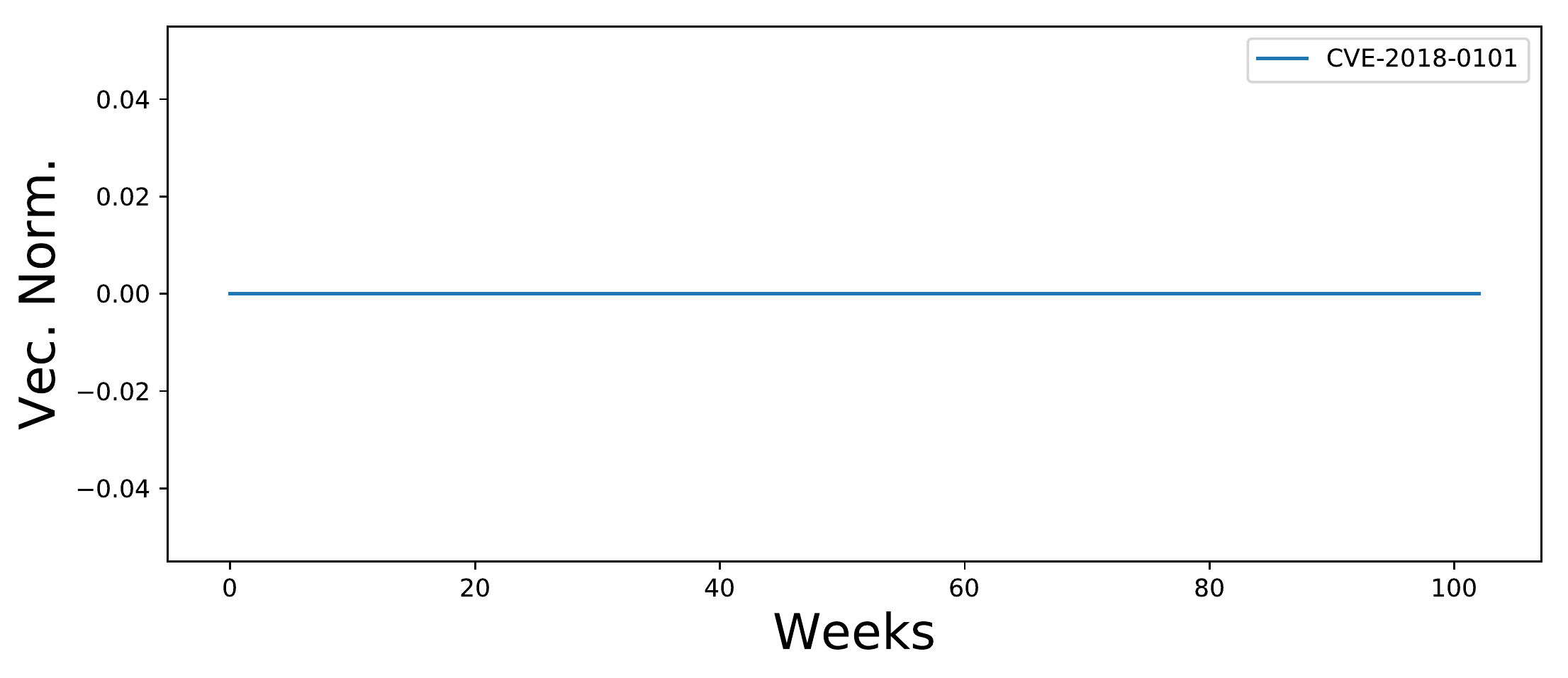}
        \caption{}
        \label{fig:popularity_e_30627}
    \end{subfigure}
	\caption{Rare event CVE-2018-0101 (Cisco Adaptive Security Appliance (ASA) RCE vulnerability). }
	\label{fig:30627} 
	\vspace{-0.3cm}
\end{figure*}

\noindent \textbf{Rare events.} The temporal security event embedding used in this paper builds on top of event frequency and event co-occurrence (see Section~\ref{sec:methodology}). By design the learned temporal event embeddings are biased towards word frequency per observation time. When analyzing the security events using the proposed method, we need to pay attention to the events that appear less frequently. Broadly speaking, these rare events can be grouped into two categories - i) the new security events associated with recently disclosed vulnerabilities and ii) those rarely observed in the IPS. For the new events, their corresponding disclosure dates are good indicators when interpreting the embedding results. For the events that rarely observed, frequency and popularity are two good reference points when interpreting the embedding results. Take ``CVE-2018-0101 (Cisco Adaptive Security Appliance (ASA) RCE vulnerability)'' for example (see Figure~\ref{fig:30627}). \approach faithfully identifies its changes over time (see Figure~\ref{fig:30627_stable}). However, when referring to event frequency (see Figure~\ref{fig:fig:popularity_f_30627}), we can see that these changes do not represent that the attackers are changing their attack campaign strategy.
Moreover, it is important to notice that our proposed system is robust and can correctly indicate that such event is not actively being exploited by attackers (see Figure~\ref{fig:popularity_e_30627}), and therefore the changes flagged at the previous steps are spurious.

\noindent \textbf{Distraction from attackers.} Our proposed temporal event embedding may be subject to distraction from malicious attackers, leading to inaccurate insights. For example, attackers could generate large amounts of fake events by targeting a considerable number of machines (\eg hundreds of thousands) over a certain period of time (\eg weeks). However, we argue that this would make the attackers more visible to the security companies who could track such malicious activities and block them accordingly. Additionally, such distraction operations would not bring financial incentive to the attackers. Note that once they switch back to real campaigns, our method would faithfully capture the new trend.

\noindent \textbf{Limitations.} \approach relies on a dataset of pre-labeled security events to generate insights and understand their evolution. An inherent limitation of this type of data is that an event can be detected only if it  belongs to a known attack vector. If, for example, a new zero-day vulnerability started being exploited in the wild, this would not be reflected in the data until its signature is created. Our method is data-driven hence it can not deduce insights before an event was detected. However, such delay can be reduced since security companies typically use threat intelligence systems and employ human specialists to analyze intrusion data identifying new attack trends. We refer interested readers to Bilge \etal~\cite{bilge2012before} for a detailed study on zero-day vulnerabilities. 

\section{Related Work}
\label{sec:related_work}
\subsection{Embedding Applications in Security}

Xu \etal~\cite{xu2017neural} proposed to use network-based graph embedding to accomplish cross-platform binary code similarity detection task. The authors adopted the structure2vec approach to effectively compute embedding vectors for the control flow graph of binary functions. This allows for efficient similarity detection by comparing the embeddings for two functions. Song \etal~\cite{song2018deepmem} propose DeepMem, a graph-based deep learning approach to automatically generate abstract representations for kernel objects and recognize these objects from raw memory dumps. 
The key idea is building a memory graph and embed the graph nodes into a low-dimensional vector space using a node's actual content and the embeddings of its four kinds of neighboring nodes. These embeddings are then used as features for classification. 
Ding \etal~\cite{ding2019asm2vec} developed an assembly code representation learning model called Asm2Vec. The key idea is to encode assembly code syntax and control flow graph into a feature vector. At the query/estimation stage, the previously unknown assembly code is encoded into a lower-dimensional vector and compared using cosine similarity. Li \etal~\cite{li2016data} introduced a data poisoning attack on matrix factorization. The authors demonstrated that, with the full knowledge of the learner, several attacks can be achieved. 

\subsection{Other Related Work}

\noindent \textbf{Concept drift.} Concept drift refers to the phenomenon that the statistical properties of the target variable change over time. Such causes less accurate predictions across time. Within the context of security research, Maggi \etal~\cite{maggi2009protecting} addressed concept drift in Web application security, while Kantchelian \etal~\cite{kantchelian2013approaches} discussed adversarial drift.  
In recent years, Jordaney \etal~\cite{jordaney2017transcend} proposed Transcend, a statistical framework to identify aging classification models. The authors used a statistical comparison of samples seen during deployment with those used to train the model, thereby building metrics for prediction quality. They then combine both decision assessment (\ie the robustness of the prediction results) and alpha assessment (\ie the quality of the non-conformity measure) to detect concept drift. 

\noindent \textbf{Empirical studies on cyberattacks.} 
Bilge \etal~\cite{bilge2012before} conducted a systematic study of the characteristics of zero-day attacks through the data collected from 11 million endpoints. Nappa \etal~\cite{nappa2015attack} conducted a systematic analysis of the patching process of 1,593 vulnerabilities in 10 client-side applications over 5 year time, especially on measuring the patching delay and several patch deployment milestones for each vulnerability. Nayak \etal~\cite{nayak2014some} carried an empirical study on vulnerability using field data and proposed several count-based metrics for attack surface evaluation. Vervier \etal~\cite{vervier2015mind} analyzed 18 months of data collected by an infrastructure specifically built to address BGP hijacks. The author characterized the BGP hijacks in this longitudinal study and provide a thorough investigation and validation of the candidate malicious BGP hijacks. Li \etal~\cite{li2017large} conducted a large-scale empirical study of security patches that affected 862 open-source projects. 

\noindent \textbf{Vulnerability prediction.} Vulnerability prediction techniques learn the attack history from previous events (e.g., historical compromise data) and use the acquired knowledge to predict future ones. What learned in the history can offer insights to evolution and is therefore relevant to our work. Sabottke \etal~\cite{sabottke2015vulnerability} conducted a quantitative and qualitative exploration of the vulnerability-related information disseminated on Twitter. The authors built a twitter-based exploit detector, which was capable of providing early warnings for the existence of real-world exploits. Similarly, Bozorgi \etal~\cite{bozorgi2010beyond} showed how to train linear support vector machines (SVMs) that predict whether and how soon a vulnerability is likely to be exploited (i.e., predict time to exploit). Recently, Shen \etal~\cite{shen2018tiresias} demonstrated that recurrent neural networks (RNNs) can be leveraged to predict the specific steps (i.e., vulnerability that may be exploited) that would be taken by an adversary when performing an attack. Liu \etal~\cite{liu2015cloudy} explored the effectiveness of forecasting security incidents. This study collected 258 externally measurable features about an organization's network covering two main categories: mismanagement symptoms (e.g., misconfigured DNS) and malicious activities (e.g., spam, scanning activities originated from this organization's network). Based on the data, the study trained and tested a Random Forest classifier on these features, and are able to achieve with 90\% True Positive (TP) rate, 10\% False Positive (FP) rate and an overall accuracy of 90\% in forecasting security incidents. In summary, these approaches learn the attack history from previous events (e.g., historical compromise data) and use the acquired knowledge to predict future ones. They don't provide thorough investigations on how security events evolve over time. 

\noindent \textbf{Alert correlation.} Alert correlation~\cite{cuppens2002alert,vasilomanolakis2015taxonomy} refers to a process that analyzes the alert logs produced by IDS and forms higher-level information on attempted intrusions. Once alerts are correlated among multiple monitors, the results can provide IDS a holistic view of the network monitored. A lot of work has been done in this areas such CRIM~\cite{cuppens2002alert}, DIDMA~\cite{kannadiga2005didma}, ACARM~\cite{valeur2004comprehensive}, INDRA~\cite{janakiraman2003indra}, etc. Vasilomanolakis \etal~\cite{vasilomanolakis2015taxonomy} summarized the current state of the art in the area of distributed and collaborative intrusion detection. In contrast to this previous work, this paper focuses on understanding the emergence, the evolution, and the characteristics of attack steps in relation to the wider context in which they are exploited.

\noindent \textbf{Automated causality analysis.} Causality is an orthogonal but interesting problem relating to \approach. Hercule~\cite{pei2016hercule} uses tainted path $s$ from $t$ to model the causality. SteamSpot~\cite{manzoor2016fast} uses a new similarity function to compare graphs and builds information flow graph clusters to detect anomalies.  NoDoze~\cite{hassan2019nodoze} builds provenance graph of a given event and use a novel diffusion algorithm to efficiently propagate and aggregate the anomalous scores. HOLMES~\cite{milajerdi2019holmes} also leverages provenance graph and identify APT attacks via information flow graphs. Conversely, our goal is to provide a reliable new method for analyzing attack trends than frequency analysis.

\noindent \textbf{Closest work.} One of the closest work to this paper is DarkEmbed~\cite{tavabi2018darkembed}. It used paragraph vector to learn low dimensional distributed representations, \ie embeddings, of darkweb/deepweb discussions. These embeddings effectively captured the meaning of these discussions and their other characteristics, such as language, and indicator words. DarkEmbed then trained a classifier to recognize posts discussing vulnerabilities that would be exploited in the wild. DarkEmbed is essentially a NLP analysis. Different from DarkEmbed, our work focuses on using representation vectors to parameterize the conditional probabilities of security events in the context of other events, and study how these security events evolve from a temporal perspective. Another closest work is~\cite{mezzour2016longitudinal}. The authors carried out a longitudinal analysis of a large corpus of cyber threat descriptions. It quantifies the severity and types (\eg worms, viruses and trojans) of 12,400 threats detected by Symantec's AV and 2,700 attacks detected by Symantec's IPS. Different from~\cite{mezzour2016longitudinal}, our work focuses on how the security events evolve and monitor how security events are exploited in the wild from real-world intrusion prevention data.

\section{Conclusion}
\label{sec:conclusion}
In this paper, we showed that techniques that were developed in the area of natural language processing can be used to effectively model and monitor the evolution of cyberattacks.
To demonstrate this, we developed \approach, a tool that leverages word embeddings to understand the context in which attack steps are exploited.
We showed that \approach is effective in flagging changes in the way attacks unfold.
In future work we plan to investigate how the use of \approach could make the work of security analysts easier in studying emerging attacks.

\section*{Acknowledgments}

We wish to thank the anonymous reviewers for their feedback and our shepherd Brad Reaves for his help in improving this paper.

\bibliographystyle{plain}
\bibliography{ref}

\end{document}